\documentclass[a4paper,12pt]{article}
\usepackage[top=1.25in, bottom=1.25in, left=1.05in, right=1.05in]{geometry}
\usepackage[utf8]{inputenc}
\usepackage[T1]{fontenc}
\usepackage[english]{babel}

\usepackage{natbib}
\usepackage[section]{placeins}
\usepackage{setspace}
\usepackage{enumitem}
\setlist{noitemsep}  

\usepackage{regexpatch}

\usepackage{amscd,amsmath,amssymb,amsfonts,amsthm,bbm,bm,latexsym,mathrsfs,mathtools,bigints}
\usepackage[subnum]{cases}	
\usepackage{dsfont}
\makeatletter
\newcommand*{\distas}[1]{\mathbin{\overset{#1}{\kern\z@\sim}}}	
\makeatother
\newcommand*\abs[1]{\left|#1\right|}		
\newcommand{\norm}[1]{\left\lVert#1\right\rVert} 		

\allowdisplaybreaks

\theoremstyle{remark}
\newtheorem{rem}{Remark}
\theoremstyle{plain}

\usepackage{algorithm, algorithmic}

\RequirePackage[colorlinks=true, citecolor=blue, urlcolor=blue]{hyperref}

\usepackage{xcolor,epsfig,epstopdf,rotating,graphics,graphicx, subcaption} 
\usepackage[width=0.9\textwidth, font={footnotesize}]{caption}

\usepackage{rotating,lscape,pdflscape}
\usepackage{booktabs,longtable,float,array,multirow,colortbl,adjustbox}
\newcolumntype{C}[1]{>{\centering\arraybackslash}p{#1}}
\definecolor{dgray}{gray}{0.65}
\definecolor{lgray}{gray}{0.90}
\definecolor{bostonred}{rgb}{0.8, 0.0, 0.0}
\definecolor{candyapplered}{rgb}{1.0, 0.03, 0.0}
\definecolor{ferrarired}{rgb}{1.0, 0.11, 0.0}
\colorlet{lred}{candyapplered!30!white}
\colorlet{dred}{ferrarired!50!lred}
\colorlet{dred}{dgray!90!white}
\colorlet{lred}{lgray!40!white}

\usepackage{pifont}

\usepackage{physics}
\usepackage{comment}

\usepackage{epsfig}

\def\I {\mathbb{I}}

\def\R {\mathbb{R}}
\def\d {\mathrm{d}}
\def\bA {\mathbf{A}}
\def\bB {\mathbf{B}}
\def\bC {\mathbf{C}}
\def\bD {\mathbf{D}}
\def\bE {\mathbf{E}}

\def\bc {\mathbf{c}}
\def\bd {\mathbf{d}}
\def\be {\mathbf{e}}
\def\bf {\mathbf{f}}
\def\bg {\mathbf{g}}
\def\bh {\mathbf{h}}
\def\br {\mathbf{r}}
\def\bs {\mathbf{s}}

\def\bw {\mathbf{w}}
\def\bx {\mathbf{x}}
\def\by {\mathbf{y}}

\def\bG {\mathbf{G}}
\def\bI {\mathbf{I}}
\def\bM {\mathbf{M}}
\def\bV {\mathbf{V}}
\def\bU {\mathbf{U}}
\def\bW {\mathbf{W}}
\def\bX {\mathbf{X}}
\def\bY {\mathbf{Y}}

\def\bzero {\mathbf{0}}
\def\balpha {\boldsymbol{\alpha}}
\def\bbeta  {\boldsymbol{\beta}}

\def\bgamma {\boldsymbol{\gamma}}

\def\bmu    {\boldsymbol{\mu}}
\def\bxi    {\boldsymbol{\xi}}

\def\bXi    {\boldsymbol{\Xi}}
\def\bSigma {\boldsymbol{\Sigma}}

\def\bOmega {\boldsymbol{\Omega}}

\def\bmuo  {\overline{\boldsymbol{\mu}}}

\def\Omegau  {\underline{\smash{\boldsymbol{\Omega}}}}
\def\Omegao  {\overline{\boldsymbol{\Omega}}}
\def\Upsilonu  {\underline{\smash{\boldsymbol{\Upsilon}}}}
\def\Upsilono  {\overline{\boldsymbol{\Upsilon}}}
\def\nuu   {\underline{\smash{\nu}}}

\def\Psiu  {\underline{\smash{\boldsymbol{\Psi}}}}

\def\rmaxk {R_k} 
\def\qgk   {q_{\bgamma,k}}
\def\qg    {q_{\gamma}}
\def\rank  {\operatorname{rank}}
\def\vec   {\operatorname{vec}}

\def\nbd   {\operatorname{nbd}}

\title{\vspace{-60pt} \textbf{Bayesian Markov-Switching Partial Reduced-Rank Regression}
}

\author{
Maria F. Pintado\thanks{CUNEF Universidad, Madrid, Spain,  {\color{blue}\texttt{mf.pintado@cunef.edu}}}
\and
Matteo Iacopini\thanks{Luiss University, Rome, Italy \color{blue}\texttt{miacopini@luiss.it}}
\and
Luca Rossini\thanks{University of Milan, Italy and Fondazione Eni Enrico Mattei, \color{blue}\texttt{luca.rossini@unimi.it}}
\and
Alexander Y. Shestopaloff\thanks{Queen Mary University of London, United Kingdom and Memorial University of Newfoundland, Canada, \color{blue}\texttt{a.shestopaloff@qmul.ac.uk}}
}

\date{\today}

\begin{document}

\maketitle

\begin{abstract}
Reduced-Rank (RR) regression is a powerful dimensionality reduction technique but it overlooks any possible group configuration among the responses by assuming a low-rank structure on the entire coefficient matrix.
Moreover, the temporal change of the relations between predictors and responses in time series induce a possibly time-varying grouping structure in the responses.
%
To address these limitations, a Bayesian Markov-switching partial RR (MS-PRR) model is proposed, where the response vector is partitioned in two groups to reflect different complexity of the relationship. A \textit{simple} group assumes a low-rank linear regression, while a \textit{complex} group exploits nonparametric regression via a Gaussian Process.
Differently from traditional approaches, group assignments and rank are treated as unknown parameters to be estimated.
Then temporal persistence in the regression function is accounted for by a Markov-switching process that drives the changes in the grouping structure and model parameters over time.
Full Bayesian inference is preformed via a partially collapsed Gibbs sampler, which allows uncertainty quantification without the need for trans-dimensional moves.
Applications to two real-world macroeconomic and commodity data  demonstrate the evidence of time-varying grouping and different degrees of complexity both across states and within each state.

 \vspace*{3pt}
    \textbf{Keywords:} Dynamic group learning; Laplace approximation; Markov switching; Rank estimation; Reduced rank regression; Uncertainty quantification.
\end{abstract}

\section{Introduction}
\label{sec:intro}
The reduced-rank (RR) linear regression model \citep{anderson1951estimating,izenman1975reduced} imposes a low-rank constraint on the coefficient matrix, achieving a smaller number of relevant linear combinations of the predictor variables that explain the variation in all the responses.
Such global covariate clustering is part of the traditional RR toolbox and includes numerous variants \citep[e.g., see][]{velu1991reduced,li2019integrative,kim2023integrative}.
%
In the presence of responses that are heterogeneous in nature (e.g., because they measure very different phenomena), the \textit{global} assumption made by RR models is likely to be violated.  
To address this issue, partial RR (PRR) regression \citep{reinsel2006partially} introduces a RR structure to only a subset of the response variables.
The response variables are divided into two non-overlapping subsets: one is assumed to be driven by a small number of linear combinations of the covariates, while the other by a full-rank coefficient (sub)matrix. Instead of relying on an a priori fixed grouping structure, this partition can be determined in a data-driven way, for example by adopting a Bayesian approach, where the group memberships and rank are treated as unknown parameters to be estimated \cite{pintado2025bayesian}.

Despite relaxing the global nature of the RR framework, a PRR model still retains the possibly strong assumption of linearity, which may be particularly restrictive in many real-world data \citep[e.g.,][]{hauzenberger2025gaussian,iacopini2025static}. 
To address this issue, our proposed model generalises the PRR by adopting a nonparametric component on one subset of the response vector.
This results in a procedure that tunes the model complexity in two ways: first, each response is allocated to either a nonparametric group via a Gaussian process (GP) or to a reduced-rank group in a data-driven fashion according to the complexity of its relationship with the predictors. Then, given the classification, the GP and the reduced-rank models are estimated.

\begin{figure}[ht!]
\centering
    \begin{subfigure}[t]{0.22\textwidth}
    \centering
    \begin{tabular}{c}
        \includegraphics[width=\linewidth]{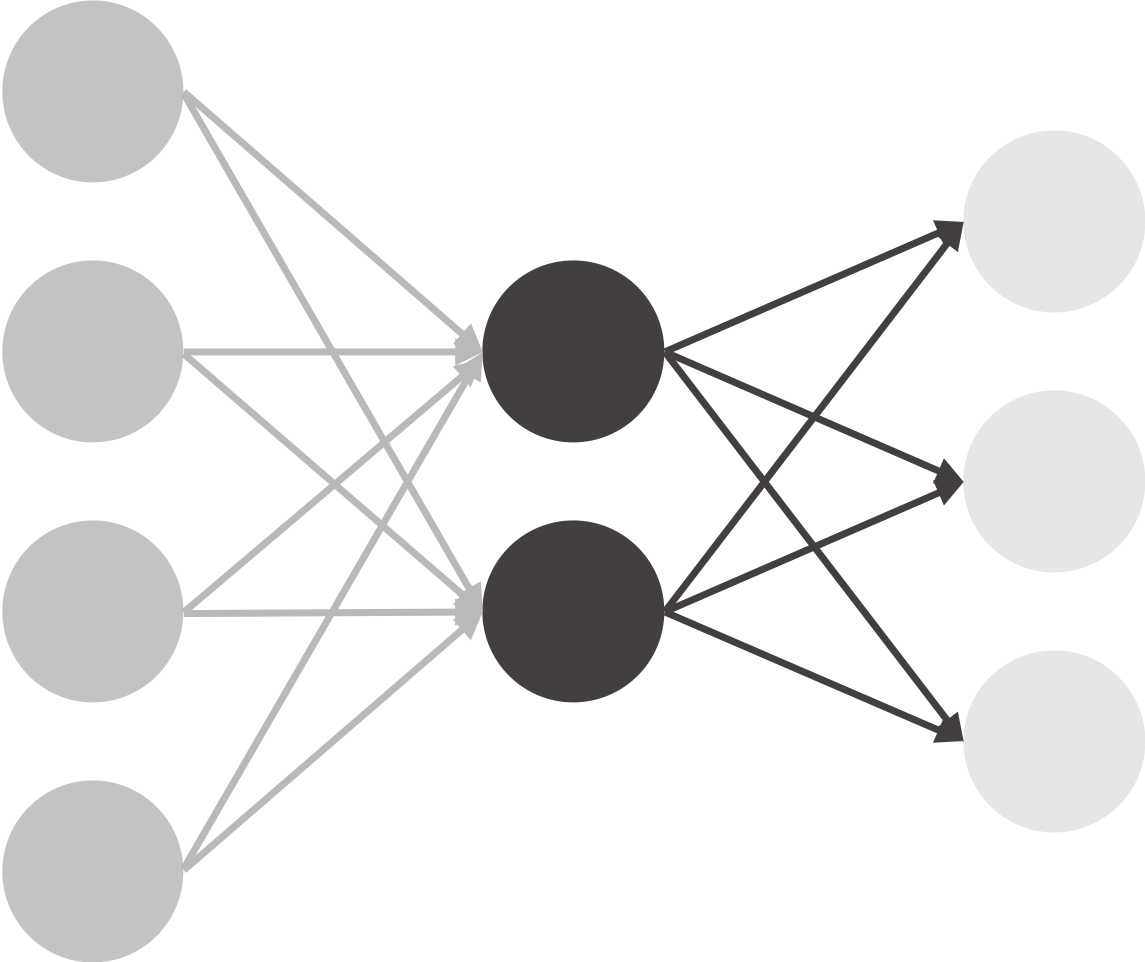}
    \end{tabular}
    \caption{RR}
    \end{subfigure}
    \hspace{13pt}
    \begin{subfigure}[t]{0.22\textwidth}
    \centering
    \begin{tabular}{c}
        \includegraphics[width=\linewidth]{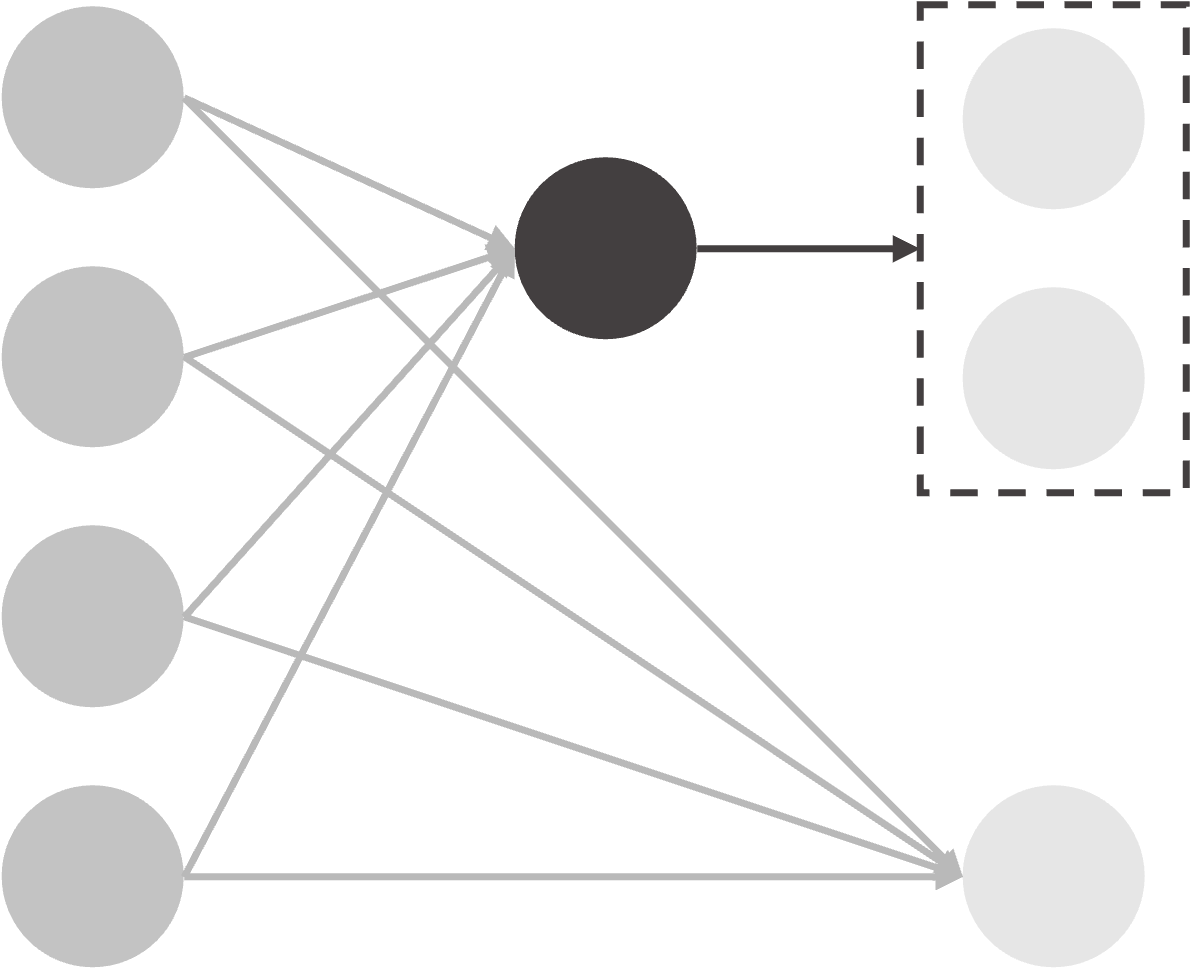}
    \end{tabular}
    \caption{PRR}
    \end{subfigure}
    \hspace{13pt}
    \begin{subfigure}[t]{0.43\textwidth}
    \centering
    \begin{tabular}{cc}
        \includegraphics[width=0.48\linewidth]{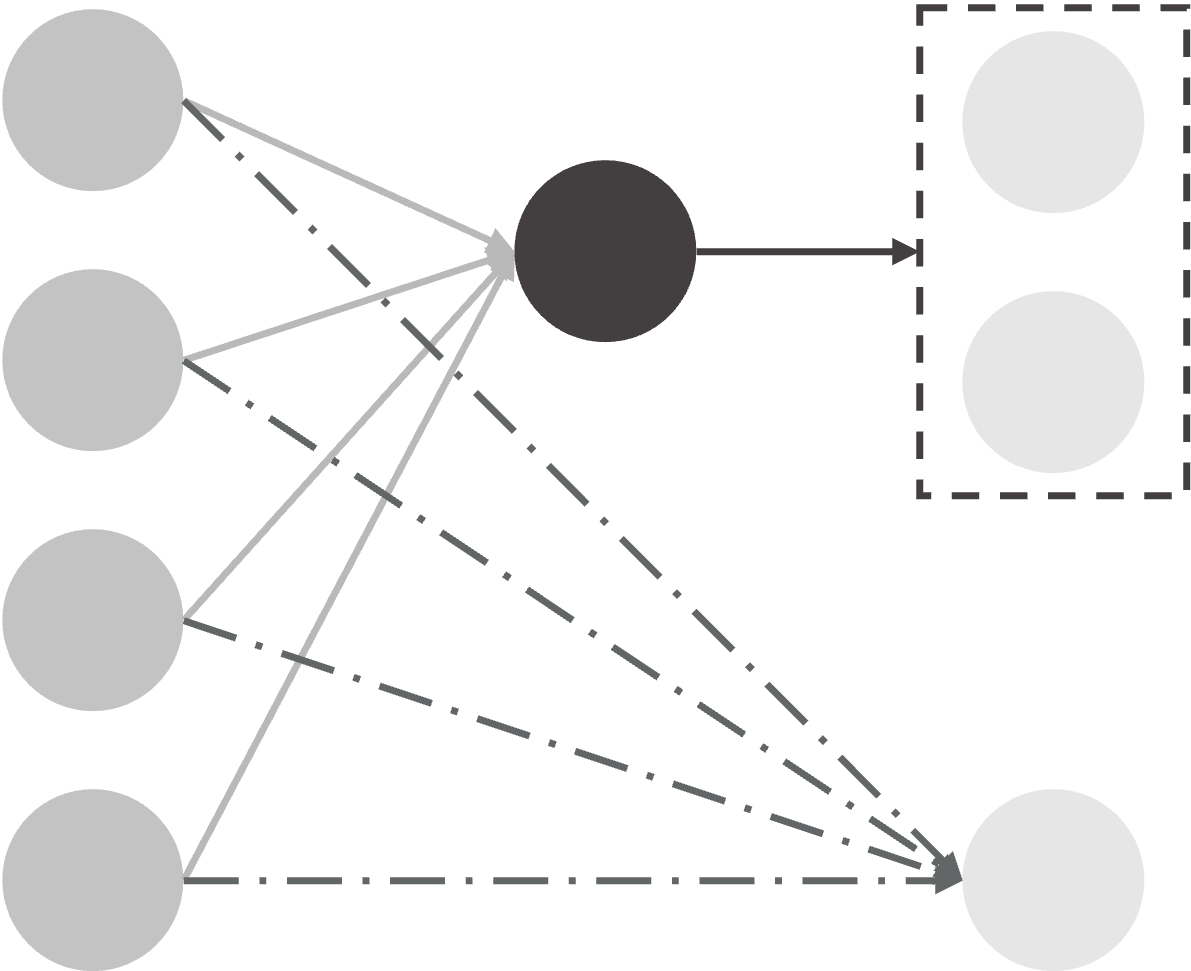} 
        & \includegraphics[width=0.48\linewidth]{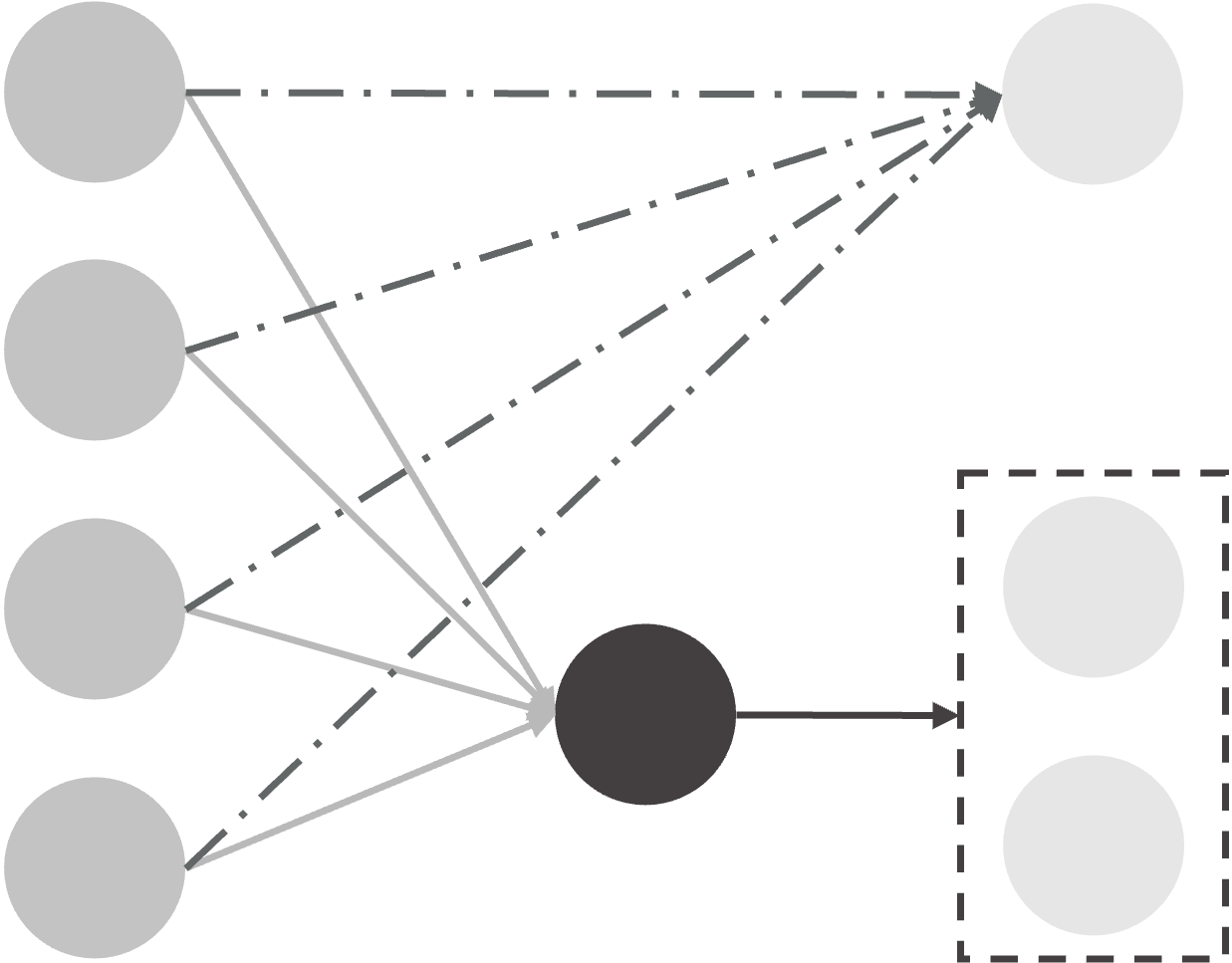} \vspace{1ex}
    \end{tabular}
    \caption{MS-PRR with two states.}
    \end{subfigure}
\caption{Interactions between four covariates (medium-grey disks) and three responses (light-grey disks), through low-rank components (dark-grey disks), and nonparametric relationships (dashes arrows) in the reduced-rank (a), partial reduced-rank (b), and Markov-switching partial reduced-rank (c) regression models.}
\label{fig:models}
\end{figure}

Notably, when multivariate data is observed through time, the grouping structure of a PRR model and the complexity of between-variable relationship can undergo changes. This has motivated the introduction of time-varying parameters in full-rank linear VAR models \citep[e.g., see][]{krolzig1997markov,primiceri2005time}.
In the context of PRR, \cite{pintado2025bayesian} recently discovered distinct complexity patterns in a macroeconomic data set by fitting a PRR model before and after a crisis.
Their finding 
leads us to extend the PRR framework further to account for time-varying response partitions using a Markov switching process, which is core for detecting period of crisis or recessions and expansions. 
Together with the GP component replacing the full-rank linear part of standard PRR, we introduce a Markov-Switching Partial Reduced-Rank (MS-PRR) regression model. Figure \ref{fig:models} illustrates the MS-PRR model compared to RR and PRR regression.
The complexity of the proposed MS-PRR model can range from most responses being described by a low-rank linear regression to being modelled by a nonparametric regression.

Our work is the first to model time-varying complexity in a PRR framework using a one-pass approach. Furthermore, our proposed approach integrates nonparametric term via GP with reduced-rank regression models, thus extending their area of applicability.
The MS-PRR is therefore able to select a very simple or very complex regression function for \textit{each response}, while also accounting for possible time-variation in the classification as well as the rank of the linear part and the degree of smoothness of the GP part.
Following the macroeconomic and commodity literature \citep{chan2018bayesian, clarkl2024forecasting}, we include in our model time-varying volatility to capture variation in the error terms and in particular, we rely on the well-known stochastic-volatility component.

In the context of time series, GPs have been adopted in the literature \citep[e.g., see][]{bonnerjee2024gaussian,clarkl2024forecasting,cunningham2012gaussian}.
On the other hand, Markov-switching models have also been implemented to allow for a time-varying and persistent clustering structure. A na\"ive approach to clustering multivariate time-series data is to apply a clustering technique independently at each time step. However, interpreting the resulting clustering structure across time would be particularly difficult and hampered by the inconsistent cluster structures over time.
Conversely, specifying a time-varying and time-dependent clustering would allow to address this issue as the structure itself would be considered persistent over time \citep[e.g., see][]{maruotti2017model,creal2014market}.

In contrast to works that focus primarily on how observations move between clusters over time \citep{paci2018dynamic,corneli2018multiple}, our approach focuses not on assigning individual units to a cluster, but on dynamically identifying persistent structural groupings within the response variables themselves, based on the complexity of their relationship with the covariates. In the MS-PRR model, we integrate PRR and hidden Markov models by introducing a Markov-switching process that drives the changes in the grouping structure and model parameters over time, accounting for time variation and persistence \citep{fruhwirth2006finite}.
Our Markov-switching component allows us to capture \textit{two} nested changes in the complexity of the predictor-response relationship. The first pertains the allocation of the different responses to either the linear or the GP component. Then, given the allocation, the complexity of the model varies across states as well. This occurs via the state-specific rank of the coefficient matrix, for the low-rank component, and hyperparameters, for the GP component.
From a computational perspective, estimating the proposed model poses nontrivial challenges as the top-level parameters and latent variables are discrete (state index and rank) and multivariate (allocation vector partitioning the responses). To tackle this issue, full Bayesian inference is performed by relying on a forward-filtering backward-sampling algorithm \citep{fruhwirth2006finite} with a carefully designed and meaningful identification restriction to prevent label switching in the estimation of the time-varying classification. Then, a partially collapsed Gibbs sampler \citep{vandyk2008partially} is used for sampling the state-specific rank, which enables us to avoid trans-dimensional samplers and enhance the mixing of the algorithm.
Finally, to sample the binary vector of allocation variables a version of the Metropolized Shotgun Stochastic Search algorithm \citep{hans2007shotgun} is implemented, which performs a local search to circumvent the computational complexity of a direct search in the entire parameter space.

In a simulation study, the proposed model is tested on various synthetic datasets to investigate its ability in recovering the time-varying response grouping structure, and the associated rank of the low-rank component, demonstrating good performance of the proposed model and inference approach.
Finally, we apply the model to two real data sets about US economic indicators and energy commodity prices, respectively. The findings evidence the persistent shifts in the response grouping in periods corresponding to the main recent global shocks (e.g., financial crisis, COVID-19 pandemic, energy crisis).

The remainder of the article is as follows.
Section~\ref{sec:model} introduces the proposed MS-PRR model and presents the prior structure. Then, Section~\ref{sec:posteriors} provides the Gibbs sampling algorithm with details from the posterior distribution.
To validate the algorithm, Section~\ref{sec:simulations} tests its performance on synthetic data, whereas Section~\ref{sec:app} provides two real-data applications to macroeconomic and commodity data.
Section~\ref{sec:conc} draws the concluding remarks.

\section{Markov-switching partial reduced-rank regression model}
\label{sec:model}

For each time point $t=1,\dots,T$, let $\by_t \in \R^q$ be the vector of response variables, and $\bx_t \in \R^p$ the vector of explanatory variables. We assume a partial reduced-rank regression model with a Markov-switching structure (MS-PRR) to account for the regime changes observable in real data, assuming the periods vary between $K$ different states driven by a hidden Markov chain with transition matrix $\bXi$. Hence, the parameters in the model are indexed by the state $k \in \{1,\dots,K\}$, while $\bs = (s_1, s_2, \dots, s_T)$ are the latent states of the Markov chain.

We assume that at each time $t$, the response vector can be partitioned into two groups of dimensions $q_{\gamma,s_t}$ and $q-q_{\gamma,s_t}$, where $q_{\gamma,s_t} \in \{ 2,\dots,q-1 \}$. The first $q_{\gamma,s_t}$ elements of $\by_t$ correspond to the first group, which admits a low-rank structure in its regression on $\bx_t$. Whilst the relationship between the following $q-q_{\gamma,s_t}$ entries and $\bx_t$ is assumed to follow a flexible model. Thus, each vector of responses is described as  $\by_t = ({y_{t,1}, \dots, y_{t,q_{\gamma,s_t}}}, {y_{t,q_{\gamma,s_t}+1}, \dots,y_{t,q}})' \in \R^q$, and an analogous structure is assumed for the error terms $\be_t \in \R^{q}$. 
We denote the matrix of regression coefficients for the low-rank component by $\bC_{s_t} \in\R^{p\times q_{\gamma,s_t}}$ with reduced rank $r_{s_t} = \rank(\bC_{s_t}) \leq \min(p,q_{\gamma,s_t})-1$, and the flexible, more complex, component modelled nonparametrically by $\bf_{s_t}(\bx_t) \in \R^{q-q_{\gamma,s_t}}$.

Therefore, the model can be represented as follows:
\begin{equation}
\label{eq:msprrmodel}
	\by_t = \bV_{1s_t}'\bC_{s_t}'\bx_t  + \bV_{2s_t}'\bf_{s_t}(\bx_t) + \be_t, \qquad \be_t \sim \mathcal{N}_q(\mathbf{0},\bSigma_t),
\end{equation}
where $\bV_{1s_t} = \left[\bI_{q_{\gamma,s_t}}, \bzero_{q_{\gamma,s_t}\times (q-q_{\gamma,s_t})} \right] \in\R^{q_{\gamma,s_t} \times q}$, and $\bV_{2s_t} = \left[\bzero_{(q-q_{\gamma,s_t}) \times q_{\gamma,s_t}},\bI_{q-q_{\gamma,s_t}}\right] \in\R^{(q-q_{\gamma,s_t}) \times q}$.
The time-varying covariance matrix assumes the decomposition $\bSigma_t = \bW^{-1} \bD_t \bW^{-1\prime}$, where $\bW$ is a lower triangular matrix with ones on the diagonal and $\bD_t = \operatorname{diag}\big( \exp(h_{1t}),\dots,\exp(h_{qt}) \big)$ with
\begin{equation}
\label{eq:SV}
    h_{jt} = h_{jt-1} + \varepsilon_{jt}, \qquad \varepsilon_{jt} \sim \mathcal{N}(0, \sigma_j^2), \qquad j=1,\dots,q,
\end{equation}
where $\sigma_j^2$ is the state variance and $h_{j0}$ is the initial condition.

\subsection{Prior specifications}
\label{sec:priors}

Following from now, we work with parameter values at a fixed state $k$. We introduce a binary vector $\bgamma_k \in \{0,1\}^q$ for each state $k=1,\dots,K$ to categorise the responses into the low-rank group and the flexible component. As we lack any prior information regarding the criteria for this classification, we assume that each element $\gamma_{j,k}$ ($j=1,\dots,q$) in each state $k=1,\dots,K$ follows an independent Bernoulli prior distribution with probability $\rho_k$ of being assigned to the low-rank group. Consequently, the joint prior distribution on $\bgamma_k$ is
\begin{equation}
\label{eq:priorgamma}
    p(\bgamma_k|\rho_k) = \left[ \prod_{j=1}^q \text{Bern}(\gamma_{j,k}|\rho_k)\right] \I\!\left( 1\!<\!q_{\bgamma,k}\!<\!q \right),
\end{equation}
where $q_{\bgamma,k}=\sum_{j=1}^{q} \gamma_{j,k}$, and $\rho_k \in (0,1)$ is the prior probability of being assigned to the low-rank group. The constraint imposed by the indicator function in Eq.~\eqref{eq:priorgamma} allows for the existence of the low-rank group and, thus, of a PRR model.
Additionally, $\rho_k$ is assigned a Beta prior distribution, $\rho_k \sim \mathcal{B}e(\rho_k | \underline{a}_{\rho},\underline{b}_{\rho})$.

The matrix of coefficients $\bC_k$ is assumed to have reduced rank $r_k \leq \rmaxk = \min(p,q_{\bgamma,k})-1$, which depends on the binary parameter $\bgamma_k$. Therefore, conditional on $q_{\bgamma,k}$ (hence on $\bgamma_k$), we assume a non-informative uniform prior distribution for $r_k$ over the discrete set $\{ 1, \dots, \rmaxk\}$, that is $r_k | \bgamma_k \sim \mathcal{U}(r_k | \{1, \dots, \rmaxk\})$.

Given that $\bC_k$ is a low-rank matrix, it can be expressed as the product of two full-rank matrices $\bA_k\in \R^{\qgk\times r_k}$ and $\bB_k\in \R^{p\times r_k}$, such that $\bC_k=\bB_k\bA_k'$. This decomposition is not unique, since for any orthogonal $r_k \times r_k$ matrix $\mathbf{P}$, we have $\bC_k = (\bB_k\mathbf{P})(\mathbf{P}'\bA_k')$. 
To achieve a unique decomposition of $\bC_k$, following \cite{geweke1996bayesian} we impose an identifying restriction by assuming the first $r_k$ rows of $\bA_k$ equal to the identity matrix $\bI_{r_k}$, 
\begin{equation}
\label{eq:A}
    \bA_k = \begin{bmatrix} \bI_{r_k} \\  \bA_{0k} \end{bmatrix},
\end{equation}
where $\bA_{0k}$ has size $(\qgk-r_k) \times r_k$. Denoting with $\vec(\cdot)$ the vectorisation operator, we assume a multivariate Gaussian prior distribution on $\balpha_{k} = \vec(\bA_{0k}')$, that is
\begin{equation}
    \balpha_{k} | \bgamma_k,r_k \sim \mathcal{N}_{(\qgk-r_k)r_k} (\balpha_{k} | \bzero,\Upsilonu_{\balpha,k}),
\end{equation}
where $\Upsilonu_{\balpha,k} = \underline{a} \, \bI_{(\qgk-r_k)r_k}$, for fixed $\underline{a} > 0$.
Similarly, defining $\bbeta_k = \vec(\bB_k)$ and letting $\Upsilonu_{\bbeta,k} = \underline{b}\, \bI_{pr_k}$, $\underline{b}>0$, we assume a multivariate Gaussian prior distribution:
\begin{equation}
    \bbeta_k | \bgamma_k,r_k \sim \mathcal{N}_{pr_k}(\bbeta_k | \bzero, \Upsilonu_{\bbeta,k}).
\end{equation}
Moving to the flexible, nonparametric component, we adopt a Gaussian process prior for the function $f_{k,j}: \R^p \to \R$, that is $f_{k,j}(\bx) \sim \mathcal{GP}\big( \underline{m}_k(\bx), \Omegau_k(\bx,\bx^*) \big)$, $j=1,\dots,q-\qgk$. Specifically, we assume zero mean $\underline{m}_k(\bx) = \mathbf{0}$ and a covariance function from the Matérn class.
We follow the conventional practice of fixing the parameter $\nu$, which controls the smoothness level of the function, as a half-integer, adhering to the popular choice $\nu=3/2$ \citep[][ch.4]{rasmussen2005gaussian}.\footnote{The covariance function from the Matérn class is given by 
$ \Omega(\bx,\bx^*) 
= \sigma_f^2 \, 2^{1-\nu}/\Gamma(\nu) 
\left( \sqrt{2\nu}\norm{\bx-\bx^*}_2/\zeta\right)^\nu
H_\nu 
\left( \sqrt{2\nu}\norm{\bx-\bx^*}_2/ \zeta\right)$,
where $\Gamma(\cdot)$ is the Gamma function, $H_\nu(\cdot)$ is a modified Bessel function, and $\nu$, $\zeta$ and $\sigma_f^2$ are positive parameters.}
This corresponds to the finite-dimensional prior $\bf_{k,j} \sim \mathcal{N}_{T_k}(\mathbf{0}, \Omegau_k)$, where the prior covariance matrix has generic element
\begin{equation}
\label{eq:kernel}
    \Omegau_{k;i,l} = \sigma_{f,k}^2 \left( 1 + \frac{\sqrt{3}\norm{\bx_i-\bx_l}_2}{\zeta_k} \right)
    \exp \left(-\frac{\sqrt{3}\norm{\bx_i-\bx_l}_2}{\zeta_k} \right),
\end{equation}
for time points $i$ and $l$ such that $s_i = s_l = k$. 

The hyperparameters, $\sigma_{f,k}^2$ and $\zeta_k$, of the Gaussian process prior control the behaviour of the function $\bf_{k,j}$ and are state dependent. 
The signal variance $\sigma_{f,k}^2$ defines the amplitude of the function, thus the variation of the function values from the mean.
The length scale $\zeta_k$ determines how quickly the process varies as a function of the input points, affecting the smoothness of the function. Hence, setting $\zeta_k$ too large yields a model
which might miss higher frequency information, whereas a $\zeta_k$ set too small leads to overfitting. 
We assign independent Gamma priors to the two hyperparameters, as follows:
\begin{equation*}
    \sigma_{f,k}^2 \sim \mathcal{G}a\big( \sigma_{f,k}^2 | \underline{a}_\sigma, \underline{b}_\sigma \big), \qquad
    \zeta_k \sim \mathcal{G}a\big( \zeta_k | \underline{a}_\zeta, \underline{b}_\zeta \big),
\end{equation*}
where $\underline{a}_\sigma, \underline{b}_\sigma, \underline{a}_\zeta,\underline{b}_\zeta$ are fixed positive values.

Regarding the covariance matrix, the free entries of $\bW$ in the covariance matrix decomposition, denoted by $\bw$, are assigned a Gaussian prior with mean equal to zero and covariance matrix equal to $\underline{\bOmega}_w$,  
while the prior imposed on $\sigma_j^2$ ($j=1,\dots,q$) is $\mathcal{IG}(\underline{a}_\sigma,\underline{b}_\sigma)$,
and $h_{j0}\sim\mathcal{N}(0,\underline{\upsilon}_j^2)$.
Lastly, we assign a Dirichlet prior distribution to each row of the transition matrix, $\bXi$, for the latent Markov chain, that is $\bxi_k \sim \mathcal{D}ir(\underline{\bd})$ for each $k=1,\dots,K$, with $\underline{\bd}\in\R_+^K$.

\subsection{Definition of the likelihood function}
\label{sec:preliminaries}

Before moving to the posterior distributions, we clarify the notation and provide further derivations for defining the likelihood function.

Let us define $\bY \in \R^{T\times q}$ as the matrix with row $t$ equal to $\by_t'$. 
Consider now a generic state $k$, and denote by $\mathcal{T}_k$ the set of time indices assigned to the $k$th state, that is $\mathcal{T}_k = \{ t\in\{1,\dots,T\} : s_t=k \}$, with cardinality $\abs{\mathcal{T}_k} = T_k$. 
Let $\widetilde{\bY}_k \in \R^{T_k \times q}$ denote the submatrix of $\bY$ containing only the rows with time index $t\in \mathcal{T}_k$, and $\widetilde{\bX}_k \in \R^{T_k\times p}$ denotes the matrix with row $t \in \mathcal{T}_k$ as $\bx_t'$.
Let $\tilde{\by}_k \in \R^{qT_k}$ denote the vector stacking the observations $\by_t$ for all the periods when $s_t=k$, that is $\tilde{\by}_k = \vec(\widetilde{\bY}_k)$, and define analogously $\tilde{\be}_k \in \R^{qT_k}$. 
In addition, $\bSigma = \{ \bW^{-1} \bD_t \bW^{-1\prime} \}_{t=1}^T$ is the collection of all the time-varying covariance matrices and similarly $\bA,\bB,\bf, \bgamma, \br$ gather the respective variables across $k = 1,\dots,K$.

Define the index set assembling the indices of all response variables allocated to the flexible group in state $k$ as $Q_k = \{ j \in \{1,\dots,q\}: \gamma_{j,k} = 0 \}$ with cardinality $q-\qgk$. 
Let $\bf = \{ f_{k,j}(\bx_t)\in \R : k=1,\dots,K, \; j\in Q_k, \; t\in \mathcal{T}_k \}$ denote the collection of functions $f_{k,j}$ evaluated at each observed covariate.
Moreover, $\bf_k = \bf_k(\bx_t) = (f_{k,1}(\bx_t),\dots,f_{k,q-\qgk}(\bx_t))' \in\R^{q-\qgk}$ and $\tilde{\bf}_k \in \R^{(q-\qgk)T_k}$ is the vector in stacking the values of $f_{k,j}$ evaluated at the observed covariate values, for all periods when $s_t=k$ and all responses belonging to the flexible group $j\in Q_k$.
%

The conditional likelihood can be rewritten as follows
\begin{align}
    p(\bY | \bs,\bSigma, \bA,\bB,\bf, \bgamma, \br) & = \prod_{k=1}^K \prod_{t \in \mathcal{T}_k} p(\by_t | s_t=k,\bSigma_t, \bA_k,\bB_k,\bf_k,\bgamma_k,r_k).
\end{align}
Parting from Eq. \eqref{eq:msprrmodel}, a vectorised model for all time periods where $s_t = k$ is given by
\begin{align}
    \tilde{\by}_k = \bU_{1k} \bc_k + \bU_{2k} \tilde{\bf}_k + \tilde{\be}_k, \qquad \tilde{\be}_k \sim \mathcal{N}_{qT_k}(\mathbf{0}, \widetilde{\bSigma}_k),
\label{eq:model_k_vectorised}
\end{align}
where $\bU_{1k} = \bV_{1k}' \otimes \widetilde{\bX}_k$, $\bc_k = \vec(\bC_k)$, $\bU_{2k} = \bV_{2k}' \otimes \bI_{T_k}$ and $\widetilde{\bSigma}_k$ is a $qT_k\times qT_k$ matrix such that $\widetilde{\bSigma}_k(t+(i-1)T_k, t+(j-1)T_k) = \bSigma_t(i,j)$, for $t=1,\dots,T_k$ and $i,j=1,\dots,q$, and has zeros in the remaining entries.
Thus, it follows $\tilde{\by}_k | \bs,\tilde{\bSigma}_k,\bA_k,\bB_k,\tilde{\bf}_k,\bgamma_k,r_k \sim \mathcal{N}_{qT_k}(\bU_{1k} \bc_k + \bU_{2k} \tilde{\bf}_k, \, \widetilde{\bSigma}_k)$.

Recall that the nonparametric prior on the scalar-valued functions $f_{k,j}(\bx) \sim \mathcal{GP}(\underline{m}_k(\bx),$ $\Omegau_k(\bx,\bx^*))$ corresponds to the finite-dimensional prior $\bf_{k,j} \sim \mathcal{N}_{T_k}(\mathbf{0}, \Omegau_k)$, where the prior covariance matrix has generic element $\Omegau_{k;i,l}$ defined in Eq. \eqref{eq:kernel}.
This implies that the vector $\tilde{\bf}_k$ has prior distribution
\begin{equation*}
    \tilde{\bf}_k \sim \mathcal{N}_{(q-\qgk)T_k}(\mathbf{0}, \Omegao_k),
\end{equation*}
where $\Omegao_k = \bI_{q-\qgk} \otimes \Omegau_k$.
Note that the conditional likelihood can be marginalized over $\tilde{\bf}_k$ analytically to obtain
\begin{align}
    p(\tilde{\by}_k | \bs,\tilde{\bSigma}_k,\bA_k,\bB_k,\bgamma_k,r_k) &= \int p(\tilde{\by}_k | \bs,\widetilde{\bSigma}_k,\bA_k,\bB_k,\tilde{\bf}_k,\bgamma_k,r_k) \, p(\tilde{\bf}_k | \bs,\bgamma_k) \, d\tilde{\bf}_k \notag \\
    &= \int \mathcal{N}_{qT_k}(\tilde{\by}_k | \bU_{1k} \bc_k + \bU_{2k} \tilde{\bf}_k, \, \widetilde{\bSigma}_k) \,
    \mathcal{N}_{(q-\qgk)T_k}(\tilde{\bf}_k | \mathbf{0}, \Omegao_k) \, d\tilde{\bf}_k \notag \\
    &= \mathcal{N}_{qT_k}(\tilde{\by}_k | \bU_{1k} \bc_k, \bSigma_{\by k}), \label{eq:marglike}
\end{align}
where $\bSigma_{\by k} = \widetilde{\bSigma}_k + \bU_{2k}\Omegao_k\bU_{2k}'$.


\section{Posterior sampling}
\label{sec:posteriors}

In this section, we design a Markov Chain Monte Carlo (MCMC) algorithm to draw samples from the joint posterior and we provide the full posterior distributions for the parameters of interest, while the derivation for the common parameters follows the literature.

In our setting, the traditional Gibbs approach that samples from the full conditional distributions of the parameters is invalid, given that the dimensions of $\bA_k$, $\bB_k$, and $\bf_k$, for each $k=1,\dots,K$, depend on the states of $\bgamma_k$ and $r_k$. 
Although reversible jump MCMC \citep{robert1999monte} deal with cases where the parameter space dimension may change across iterations of the MCMC algorithm, its practical implementation would require the definition of cross-model moves, which are hard to define and execute properly in complex settings like ours. 

To overcome this challenge, we adapt the sampling strategy proposed in \citep{pintado2025bayesian} to our framework, and implement a partially collapsed Gibbs sampler \citep[PCG, see][]{vandyk2008partially}.
%
In particular, for each state $k$, we draw $(\bgamma_k,r_k)$ from a joint distribution marginalised over the parameters $(\bA_k,\bB_k,\bf_k)$ whose sizes depend on $(\bgamma_k,r_k)$, which obviates the need of transdimensional samplers, similar to \cite{yang2022rrr}. 
Afterwards, $(\bA_k,\bB_k,\bf_k)$ is sampled conditionally on the updated values of $(\bgamma_k,r_k)$.
The entire sampling process is summarised in Algorithm~\ref{alg:PCGSV}, while the relevant steps are provided in the following subsections.

\begin{algorithm} 
\caption{PCG for Bayesian MS-PRR model}
\label{alg:PCGSV}
\algsetup{linenodelimiter=.}
\begin{algorithmic}[1]
\FOR{$k=1,\dots,K$}
    \STATE Sample $\bgamma_k$ from $p(\bgamma_k|\widetilde{\bY}_k,\bs,\bSigma,\rho_k)$.
    \STATE Sample $r_k$ from $p(r_k|\widetilde{\bY}_k,\bs,\bSigma,\bgamma_k)$.
    \STATE Sample $\tilde{\bf}_{k}$ from $p(\tilde{\bf}_{k}|\widetilde{\bY}_k,\bs,\bSigma,\bgamma_k,\bA_k,\bB_k,\sigma_{f,k}^2,\zeta_k) = \mathcal{N}_{(q-\qgk)T_k}(\overline{\bmu}_{\bf,k}, \Upsilono_{\bf,k})$.
    \STATE Sample $\balpha_k = \vec(\bA_{0k}')$ from $p(\balpha_k|\widetilde{\bY}_k,\bs,\bSigma,\bgamma_k,r_k,\bB_k,\tilde{\bf}_k) = \mathcal{N}_{(\qgk-r_k)r_k}(\overline{\bmu}_{\alpha,k}, \Upsilono_{\alpha,k})$, then set $\bA_k = \left[\bI_{r_k}, \bA_{0k}'\right]'$.
    \STATE Sample $\bbeta_k = \vec(\bB_k)$ from $p(\bbeta_k|\widetilde{\bY}_k,\bs,\bSigma,\bgamma_k,r_k,\bA_k,\tilde{\bf}_k) = \mathcal{N}_{pr_k}(\overline{\bmu}_{\beta,k}, \Upsilono_{\beta,k})$.
    \STATE Sample $\rho_k$ from $p(\rho_k | \bgamma_k) = \mathcal{B}e(\overline{a}_\rho, \overline{b}_\rho)$.
    \STATE Sample $\bxi_k$ from $p(\bxi_k|\bs) = \mathcal{D}ir(\overline{\bd}_\xi)$.
    \STATE Sample $\zeta_k$ from $p(\zeta_k | \tilde{\bf}_{k}, \sigma_{f,k}^2, \underline{a}_\zeta,\underline{b}_\zeta)$
    \STATE Sample $\sigma_{f,k}^2$ from $p(\sigma_{f,k}^2 | \tilde{\bf}_{k}, \zeta_k, \underline{a}_\sigma, \underline{b}_\sigma)$
\ENDFOR
\STATE Sample $\bs$ from $p(\bs|\bY,\bgamma,\bA,\bB,\bf,\bSigma,\bXi)$.
\FOR{$j=1,\dots,q$}
    \STATE Sample $\bh_j$ from $p(\bh_j|\bY,\bs,\bgamma,\bA,\bB,\bf,\bW,\sigma_j^2)$, using the auxiliary mixture sampler of \cite{omori2007stochastic}.
    \STATE Sample $h_{j0}$ from $p(h_{j0} | \sigma_j^2,\bh_j) = \mathcal{N}(\overline{\mu}_h, \overline{\upsilon}_h^2)$.
    \STATE Sample $\sigma_j^2$ from $p(\sigma_j^2 | \bh_j) = \mathcal{IG}(\overline{a}_{\sigma,j}, \overline{b}_{\sigma,j})$.
\ENDFOR
\STATE Sample $\bw$ from $p(\bw | \bY,\bs,\bgamma,\bA,\bB,\bf,\bh) = \mathcal{N}_{q(q-1)/2}(\overline{\bmu}_w, \overline{\bSigma}_w)$.
\end{algorithmic}
\end{algorithm}


\subsection{Sampling the state-specific allocations and rank}
\label{sec:gammar}

We aim at sampling $\bgamma_k$ from the conditional posterior marginalised over $(r_k,\bA_k,\bB_k,\bf_k)$, that is
\begin{equation}
\label{eq:posteriorgamma}
    p(\bgamma_k|\tilde{\by}_k,\bs,\widetilde{\bSigma}_k,\rho_k) = \frac{p_{\gamma_k}(\tilde{\by}_k|\bs,\widetilde{\bSigma}_k,\bgamma_k) p(\bgamma_k|\rho_k)}{\sum_{\bgamma_k^\dag\in\{0,1\}^q} p_{\gamma_k}(\tilde{\by}_k|\bs,\widetilde{\bSigma}_k,\bgamma_k^{\dag}) p(\bgamma_k^{\dag}|\rho_k)},
\end{equation}
where $p_{\gamma_k}(\tilde{\by}_k|\bs,\tilde{\bSigma}_k,\bgamma_k)$ is defined as
\begin{align}
\label{eq:likelihoodgamma}
    p_{\gamma_k}(\tilde{\by}_k | \bs,\tilde{\bSigma}_k,\bgamma_k) & = \!\sum_{r_k=1}^{\rmaxk} \frac{1}{\rmaxk} \!\iint p(\tilde{\by}_k | \bs,\tilde{\bSigma}_k,\bA_k,\bB_k,\bgamma_k,r_k) \, p(\bA_k,\bB_k | \bgamma_k,r_k) \, \d\bA_k \, \d\bB_k \\
    \notag
    & = \sum_{r_k=1}^{\rmaxk} \frac{1}{\rmaxk} p_{r_k}(\tilde{\by}_k | \bs,\tilde{\bSigma}_k,\bgamma_k,r_k).
\end{align}
However, the integral of $\bA_k,\bB_k$ in Eq.~\eqref{eq:likelihoodgamma} is analytically intractable, thus we rely on the Laplace method \citep{raftery1995bayesian} to approximate it and we obtain
\begin{equation}
\label{eq:loglikelihoodgammar}
    \log p_{r_k}(\tilde{\by}_k | \bs,\widetilde{\bSigma}_k,\bgamma_k,r_k) \approx \log p(\tilde{\by}_k | \bs,\widetilde{\bSigma}_k,\hat{\bA}_k,\hat{\bB}_k,\bgamma_k,r_k) -\frac{p r_k+(\qgk-r_k)r_k}{2} \log T_k,
\end{equation}
where $\widehat{\bA}_k$ and $\widehat{\bB}_k$ are the maximum likelihood estimators (MLEs) of $\bA_k$ and $\bB_k$ given $(\bgamma_k,r_k)$. 
Deriving these MLEs involves non-trivial steps arising from two main aspects. 
First, the covariance matrix of the Gaussian density in Eq. \eqref{eq:marglike} incorporates heteroscedastic errors through the dependence of $\bU_{2k}$ on $\bV_{2k}$. 
Second, two structural restrictions are imposed on the model: the identification restriction on the matrix $\bA_k$, and the representation of the response vector as the sum of the low-rank and full-rank components in Eq. \eqref{eq:msprrmodel} through the binary matrices $\bV_{1k}$ and $\bV_{2k}$. We exploit the general class of reduced-rank regression models (GRRR) studied in \cite{hansen2002generalized}, where a ML estimation technique able to accommodate our setting is proposed.

The GRRR problem considers the regression for all periods when $s_t=k$
\begin{equation}
    \by_t = \bV_{1k}'\bA_k\bB_k'\bx_t' + \bar{\be}_t, \quad \text{ for } t\in\mathcal{T}_k,
\end{equation}
where $\bar{\be}_t$ is column $t$ of $\bar{\bE}\in \R^{q \times T_k}$, and $\vec(\bar{\bE}) \sim \mathcal{N}_{qT_k}(\bzero,\bSigma_{\by k})$, subject to the restriction
\begin{equation*}
    \vec(\bV_{1k}'\bA_k) = \bG_k \vec(\bA_{0k}) + \bg_k,
\end{equation*}
where $\bG_k$ is a binary matrix of dimension $qr_k \times r_k(\qgk - r_k)$ and $\bg_k$ is the $qr_k$-dimensional vector of restrictions. 
In detail, $\bg_k$ has entries with ones in the index set $\{(\ell-1)(q+1)+1, \; \text{ for each } \ell = 1,2,\dots,r_k\}$, and zeros in the remaining entries. The matrix $\bG_k$ is a block matrix defined as
\begin{align*}
    \bG_k       &= [\bG_{k1}, \bG_{k2}, \dots, \bG_{k r_k}], \\
    \bG_{k\ell} &= [\bzero_{(\qgk-r_k) \times (q(\ell-1)+r_k)}, 
           \bI_{\qgk-r_k}, 
           \bzero_{(\qgk-r_k)\times(q(r_k-\ell+1)-\qgk)}]',
           \qquad \ell = 1,2, \dots, r_k.
\end{align*}


Notice that there are no constraints on the matrix $\bB_k$. Then, the MLEs of $\balpha_{\bV_{1k}} = \vec(\bV_{1k}\bA_k)$ and $\bbeta_k = \vec(\bB_k)$ following the GRRR method are obtained as
\begin{align}
\label{eq:mlesa}
    \hat{\balpha}_{\bV_{1k}} &= \bG_k (\bG_k'\bM_{\bB_k} \bG_k)^{-1} \bG_k'(\mathbf{n}_{\bB_k}-\bM_{\bB_k} \mathbf{g}_k) + \mathbf{g}_k,\\
\label{eq:mlesb}
    \hat{\bbeta}_k &= \bM_{\bA_k}^{-1} \mathbf{n}_{\bA_k},
\end{align}
where 
$\bM_{\bB_k} = (\widetilde{\bX}_k\bB_k \otimes \bI_q)' \widetilde{\bSigma}_{\by k}^{-1} (\widetilde{\bX}_k\bB_k \otimes \bI_q)$,
$\mathbf{n}_{\bB_k} = (\widetilde{\bX}_k\bB_k \otimes \bI_q)' \widetilde{\bSigma}_{\by k}^{-1} \vec(\widetilde{\bY}_k')$,
$\bM_{\bA_k} = \mathbf{K}_{p,r_k}' (\widetilde{\bX}_k \otimes \bV_{1k}'\bA_k)' \widetilde{\bSigma}_{\by k}^{-1} (\widetilde{\bX}_k \otimes \bV_{1k}'\bA_k) \mathbf{K}_{p,r_k}$,
$\mathbf{n}_{\bA_k} = \mathbf{K}_{p,r_k}' (\widetilde{\bX}_k \otimes \bV_{1k}'\bA_k)' \widetilde{\bSigma}_{\by k}^{-1} \vec(\widetilde{\bY}_k')$,
$\widetilde{\bSigma}_{\by k} = \mathbf{K}_{T_k,q} \bSigma_{\by k} \mathbf{K}_{T_k,q}'$,
and $\mathbf{K}_{m,n}$ is the $mn \times mn$ commutation matrix, which transforms the vectorisation of a matrix $\bM \in \mathbb{R}^{m\times n}$ into the vectorisation of its transpose, such that $\mathbf{K}_{m,n} \vec(\bM) = \vec(\bM')$.

Since the MLEs in Eq.~\eqref{eq:mlesa} and \eqref{eq:mlesb} are cross-dependent, the practical implementation of the GRRR method is done recursively, updating $\balpha_{\bV_{1k}}$ and $\bbeta_k$ iteratively until convergence.
%
Once a solution $(\hat{\balpha}_{\bV_{1k}},\hat{\bbeta}_k)$ is obtained, it suffices to transform the vectorised MLEs to their matrix forms $\bV_{1k} \widehat{\bA}_k'$ and $\widehat{\bB}_k$ to obtain the MLEs of the low-rank coefficient matrix as $\widehat{\bC}_k = \widehat{\bB}_k (\bV_{1k}')^+ \bV_{1k}' \widehat{\bA}_k = \widehat{\bB}_k \widehat{\bA}_k'$, where $(\bV_{1k}')^+$ denotes the Moore-Penrose pseudoinverse of $\bV_{1k}'$.\footnote{Note that $\bV_{1k}$ is not invertible since its dimensionality is $\qgk \times q$, with $\qgk < q$.}


We obtain the approximation
\begin{equation}
    p_{\gamma_k}(\tilde{\by}_k | \bs,\widetilde{\bSigma}_k,\bgamma_k) \approx \sum_{r_k=1}^{\rmaxk} \frac{1}{\rmaxk} \tilde{p}_{r_k}(\tilde{\by}_k | \bs,\widetilde{\bSigma}_k,\bgamma_k,r_k) \equiv \tilde{p}_{\gamma_k}(\tilde{\by}_k | \bs,\widetilde{\bSigma}_k,\bgamma_k),
\end{equation}
where $\tilde{p}_{r_k}(\tilde{\by}_k | \bs,\widetilde{\bSigma}_k,\bgamma_k,r_k)$ is the Laplace approximation of $p_{r_k}(\tilde{\by}_k | \bs,\widetilde{\bSigma}_k,\bgamma_k,r_k)$ obtained from Eq.~\eqref{eq:loglikelihoodgammar} to the integral in Eq.~\eqref{eq:likelihoodgamma}.
Therefore, the (partially marginal) posterior distribution of $\bgamma_k$ is approximated by
\begin{equation}
\label{eq:posteriorgammaapprox}
    \tilde{p}(\bgamma_k | \tilde{\by}_k,\bs,\widetilde{\bSigma}_k,\rho_k) = \frac{\tilde{p}_{\gamma_k}(\tilde{\by}_k | \bs,\widetilde{\bSigma}_k,\bgamma_k) p(\bgamma_k|\rho_k)}{\sum_{\bgamma_k^\dag\in\{0,1\}^q} \tilde{p}_{\gamma_k}(\tilde{\by}_k | \bs,\widetilde{\bSigma}_k,\bgamma_k^{\dag}) p(\bgamma_k^{\dag}|\rho_k)}.
\end{equation}
Provided that $\bgamma_k$ is a $q$-dimensional binary vector, the total number of possible configurations for the response allocation is $2^q$, which is computationally intensive even for moderate values of $q$. 
To circumvent this problem, we employ an approximate sampling strategy for $\bgamma_k$, namely the Metropolized Shotgun Stochastic Search (MSSS) algorithm introduced by \citet{hans2007shotgun}.
The MSSS algorithm efficiently explores the high-dimensional parameter space by evaluating a subset of neighbouring configurations of the current state of $\bgamma_k$. Restricting the neighbourhood to a manageable set introduces a balance between exploration of the parameter space and computational feasibility.
Following the approach of \citet{yang2022rrr}, we define the neighbourhood of $\bgamma_k$ as all binary vectors that differ from $\bgamma_k$ in exactly one component, while still satisfying the existence of a low-rank group.
%
This neighbourhood restriction improves computational efficiency compared to considering all possible configurations, while still enabling an effective (albeit reduced) exploration of the space.
A proposal distribution is defined by
\begin{equation*}
\label{eq:ggamma}
g(\bgamma_k|\bgamma_k^{(m)}) \propto \tilde{p}(\bgamma_k|\tilde{\by}_k,\bs,\widetilde{\bSigma}_k,\rho_k) \, \I \left(\bgamma_k \in \nbd(\bgamma_k^{(m)})\right),
\end{equation*}
where $\bgamma_k^{(m)}$ denotes the value of $\bgamma_k$ at the $m$th iteration of the MCMC.
Hence, for the $k$th state of the hidden Markov chain, the proposed sampling scheme draws the allocation vector from the marginal posterior $p(\bgamma_k | \tilde{\by}_k,\bs,\widetilde{\bSigma}_k,\rho_k)$ following the steps:
\begin{enumerate}[label=\arabic*.]
    \item Generate $\bgamma_k^*$ from $g(\bgamma_k|\bgamma_k^{(m)})$.
    \item Accept $\bgamma_k^{(m+1)}=\bgamma_k^*$ with probability 
        \begin{equation*}
        \label{eq:probgamma}
            \rho_{\gamma_k} = \min\left\{1,
            \frac{\sum_{\bgamma_k\in\nbd(\bgamma_k^{(m)})} \tilde{p}_{\bgamma_k}(\tilde{\by}_k|\bs,\widetilde{\bSigma}_k,\bgamma_k) p(\bgamma_k|\rho_k)}
            {\sum_{\bgamma_k^{\dag}\in\nbd(\bgamma_k^*)} \tilde{p}_{\bgamma_k^{\dag}}(\tilde{\by}_k|\bs,\widetilde{\bSigma}_k,\bgamma_k^{\dag}) p(\bgamma_k^{\dag}|\rho_k)}
            \right\},
        \end{equation*}
        otherwise set $\bgamma_k^{(m+1)}=\bgamma_k^{(m)}$.
\end{enumerate}

The conditional posterior of $r_k$, marginalised over $(\bA_k,\bB_k,\tilde{\bf}_k)$, is approximated through the Laplace method, following the same steps described for $\bgamma_k$. Specifically, we compute the approximated (partially marginal) posterior
\begin{equation}
\label{eq:posteriorrapprox}
    \tilde{p}(r_k | \tilde{\by}_k,\bs,\widetilde{\bSigma}_k,\bgamma_k) = \frac{\tilde{p}_{r_k}(\tilde{\by}_k | \bs,\widetilde{\bSigma}_k,\bgamma_k,r_k) p(r_k|\bgamma_k)}{\sum_{r_k^{\dag}=1}^{\rmaxk} \tilde{p}_{r_k^{\dag}}(\tilde{\by}_k | \bs,\widetilde{\bSigma}_k,\bgamma_k, r_k^{\dag}) p(r_k^{\dag}|\bgamma_k)},
\end{equation}
where $\tilde{p}_{r_k}(\tilde{\by}_k | \bs,\widetilde{\bSigma}_k,\bgamma_k,r_k) p(r_k|\bgamma_k)$ is obtained in Eq.~\eqref{eq:loglikelihoodgammar}.
A new value of $r_k$ is sampled from the discrete distribution on $\{ 1,\dots,\rmaxk\}$ with the probabilities given in Eq.~\eqref{eq:posteriorrapprox}.
This process is iterated for each state $k=1,\dots,K$ to sample the collections $\bgamma = \{\bgamma_1,\dots,\bgamma_K \}$ and $\br = \{r_1,\dots,r_K \}$.

\subsection{Sampling the nonparametric term}
\label{sec:nonparametric}

Using the function space approach, we derive the posterior distribution of the nonparametric element, $\tilde{\bf}_k$. The distribution of the vectorised model in Eq. \eqref{eq:model_k_vectorised} implies that the vector collecting all the partial residuals from observations allocated to state $k$, $\overline{\by}_{1k} = \tilde{\by}_k - \bU_{1k} \bc_k$, is distributed as
\begin{equation}
\label{eq:ybar_1k}
    \overline{\by}_{1k} | \widetilde{\bSigma}_k,\tilde{\bf}_k \sim \mathcal{N}_{qT_k}(\bU_{2k} \tilde{\bf}_k, \, \widetilde{\bSigma}_k).
\end{equation}
Notice that the dependence on $(\bs,\bA_k,\bB_k)$ is implicitly accounted for in the definition of $\overline{\by}_{1k}$. 
Indeed, we face a particular challenge in the proposed PCG sampler associated with the dimensions of $(\bA_k,\bB_k,\tilde{\bf}_k)$, which do not necessarily agree between subsequent MCMC iterations, given their dependence on $(\bgamma_k, r_k)$. 
Accordingly, we overcome this issue in the current Step 4 of Algorithm \ref{alg:PCGSV} by defining an auxiliary matrix $\bC_{k}^*$ made of the first (newly sampled value) $\qgk$ columns of the matrix $\bC_k$ updated in the previous MCMC iteration \citep{pintado2025bayesian,yang2022rrr}. 
Hence, the vector $\overline{\by}_{1k}^* = \tilde{\by}_k - \bU_{1k} \bc_{k}^*$, where $\bc_{k}^*=\vec(\bC_{k}^*)$ maintains the distribution in Eq. \eqref{eq:ybar_1k}, and it is used 
for the likelihood of $\tilde{\bf}_k$.



Therefore, the posterior distribution of $\tilde{\bf}_k$ can be obtained by applying Bayes' rule, resulting in
\begin{align*}
    p(\tilde{\bf}_k | \widetilde{\bY}_k,\bs,\bSigma,\bA_k,\bB_k,\bgamma_k,\sigma_{f,k}^2,\zeta_k) & \propto \mathcal{N}_{(q-\qgk)T_k}(\tilde{\bf}_k | \mathbf{0}, \Omegao_k) \mathcal{N}_{qT_k}(\overline{\by}_{1k}^* | \bU_{2k} \tilde{\bf}_k, \widetilde{\bSigma}_k) \\
    & \propto \mathcal{N}_{(q-\qgk)T_k}(\tilde{\bf}_k | \bmuo_{\bf,k}, \Upsilono_{\bf,k}),
\end{align*}
with $\Upsilono_{\bf,k} = \big( \Omegao_k^{-1} + \bU'_{2k} \widetilde{\bSigma}_k^{-1} \bU_{2k} \big)^{-1}$ and 
$\bmuo_{\bf,k} = \Upsilono_{\bf,k}  \bU_{2k}'\widetilde{\bSigma}_k^{-1} \overline{\by}_{1k}^*$.

Regarding the GP hyperparameters, we approximate the posterior full conditional distributions of $\sigma_{f,k}^2$ and $\zeta_k$ using a Griddy Gibbs approach \citep{ritter1992facilitating}, which discretises the space over a fixed grid of length $d_u$ equal to $100$ for each parameter.

First, recall from Section \ref{sec:preliminaries} that the vector $\bf_{k,j} \in \R^{T_k}$ has a normal distribution with zero mean and covariance $\Omegau_k$.
Then, recall from Eq. \eqref{eq:kernel} that the kernel $\Omegau_k \in \R^{T_k\times T_k}$ depends on the values of the hyperparmeters, and we rewrite as $\Omegau_{k}(\sigma_{f,k}^2,\zeta_k)$. 
The full conditional distributions are
\begin{align*}
    \zeta_k | \bullet & \propto p\big( \zeta_k \big) \times \prod_{j \in Q_k} \abs{\Omegau_{k}(\sigma_{f,k}^2,\zeta_k)}^{-\frac{1}{2}} 
    \exp\Big\{ -\frac{1}{2} \bf_{k,j}' \Omegau_{k}(\sigma_{f,k}^2,\zeta_k)^{-1} \bf_{k,j} \Big\}, \\
    \sigma_{f,k}^2 | \bullet & \propto p \big( \sigma_{f,k}^2 \big)
    \times \prod_{j \in Q_k} \abs{\Omegau_{k}(\sigma_{f,k}^2,\zeta_k)}^{-\frac{1}{2}} 
    \exp\Big\{ -\frac{1}{2} \bf_{k,j}' \Omegau_{k}(\sigma_{f,k}^2,\zeta_k)^{-1} \bf_{k,j} \Big\}.
\label{eq:z_sig2f_posteriors}
\end{align*}

\begin{rem}
The vector $\tilde{\bf}_k$ contains the GP component of the regression for all the $q-\qgk$ variables allocated to the flexible group. Therefore, by drawing $\tilde{\bf}_k$ we are jointly sampling all the vectors $\bf_{k,1},\dots,\bf_{k,q-\qgk}$.
Since the posterior values of the hyperparameters involve computations with several sparse matrices, sampling from the $(q-\qgk)T_k$-dimensional Gaussian can be made efficiently.
\end{rem}

\subsection{Sampling the latent states}
\label{sec:states}

An efficient approach to sampling the latent state chain in Markov switching models is multi-move sampling, which involves drawing the entire sequence of states, $\bs=(s_1,\dots,s_T)$, jointly from its conditional posterior distribution. We employ a forward-filtering-backward sampling algorithm \citep[e.g., see][chap.~11]{fruhwirth2006finite} to sample a path of the hidden Markov chain. The detailed steps are described in the Supplement.

As known in the literature, Markov-switching models suffer from a label-switching problem, an identification issue due to invariance to permutations of the regime labels. A standard approach to address this challenge is to impose identification constraints.

\begin{rem}[Identification constraint]
The states are identified by assigning labels in correspondence to the number of observations allocated to each regime, with $k=1$ assigned to the regime with the largest observation count, such that $T_1 > T_2 > \dots > T_K$. In the case where two or more states have the same number of time periods, we assign the labels in inverse correspondence to the number of low-rank responses:  $q_{\gamma,1} > q_{\gamma,2} > \dots > q_{\gamma,K}$. However, if these values coincide, then we look at the distance between the current vector $\bgamma^{(m)}$ and the one at the previous iteration $\bgamma^{(m-1)}$. For the case where $K=2$, this step is as follows:

\begin{enumerate}
    \item Construct a $2\times 2$ matrix $\bD$ with entries
    \begin{align*}
        D_{11} &= \sum_{j=1}^q |\gamma_{j,1}^{(m)} -\gamma_{j,1}^{(m-1)}|, \qquad
        D_{12} = \sum_{j=1}^q |\gamma_{j,1}^{(m)} -\gamma_{j,2}^{(m-1)}|, \\
        D_{21} &= \sum_{j=1}^q |\gamma_{j,2}^{(m)} -\gamma_{j,1}^{(m-1)}|, \qquad
        D_{22} = \sum_{j=1}^q |\gamma_{j,2}^{(m)} -\gamma_{j,2}^{(m-1)}|.
    \end{align*}

    \item Construct the $2\times 2$ matrix $\bD^{norm} = \frac{\exp(-D_{uv})}{\sum_{w=1}^{2} \exp(-D_{uw})}$, and compute $D_{val} = D_{12}^{norm} + D_{21}^{norm} - (D_{11}^{norm} + D_{22}^{norm})$.
    \begin{enumerate}
        \item if $D_{val} \leq d_{val}$, where $d_{val}$ is an arbitrary threshold, then keep the labels from the previous iteration.
        \item if $D_{val} > d_{val}$, then exchange the labels.
    \end{enumerate}
\end{enumerate}

When $K>2$, the previous steps are easily extended to a $K\times K$ matrix $\bD$, and choose the label permutation that minimises the total distance.
\end{rem}

\subsection{Sampling the other parameters}
\label{sec:otherparams}

Regarding the sampling of the other parameters, e.g., Steps 5-8 of Algorithm \ref{alg:PCGSV}, we describe them quickly below.

The update of $\bA_k = \left[\bI_{r_k},\bA_{0k}'\right]'$ given $(\widetilde{\bY}_k, \bs,\bSigma,\bgamma_k, r_k, \bB_k, \tilde{\bf}_k)$ is performed by sampling $\balpha_k = \vec(\bA_{0k}')$. 
We define the matrix $\bB_{k}^*$ following the same strategy as in Section \ref{sec:nonparametric} to avoid dimensional incompatibility. Recalling the identification constraint in $\bA_k$ and the factorisation $\bC_{k} = \bB_k \bA_k'$, one obtains $\bC_k = \left[\bB_k,\bB_k\bA_{0k}'\right]$. Thus, we select the first $r_k$ columns of $\bC_k$, using the newly sampled value of the rank in state $k$ at the current MCMC iteration to construct $\bB_{k}^*$. 
Therefore, the posterior distribution of $\balpha_k$ is proportional to the multivariate Gaussian distribution 
\begin{align*}
    p(\balpha_k | \bY,\bs,\bSigma,\bgamma_k,r_k,\bB_{k}^*,\bf_k) 
    & \propto p(\balpha_k|\bgamma_k,r_k) \, p(\bY|\bs,\bSigma,\bA_k,\bB_{k*},\tilde{\bf}_k) \\
    & \propto \mathcal{N}_{(\qgk-r_k)r_k}(\balpha_k | \bmuo_{\balpha,k}, \Upsilono_{\balpha,k})
\end{align*}
with mean $\bmuo_{\balpha,k} = \Upsilono_{\balpha,k}\big( \mathbf{m}_{\left[J_k\right]} - \mathbf{H}_{\left[J_k,J_k\right]} \mathbf{v}_k \big)$ 
and covariance matrix $\Upsilono_{\balpha,k} = \big(\Upsilonu_{\balpha,k}^{-1} + \mathbf{H}_{\left[J_k,J_k\right]} \big)^{-1}$, where 
$\mathbf{v}_k = \vec(\bI_{r_k})$, 
$\mathbf{m}_k = \bM_{\balpha,k}' \widetilde{\bSigma}_k^{-1} \overline{\by}_{2k}$, 
$\mathbf{H}_k = \bM_{\balpha,k}' \widetilde{\bSigma}_k^{-1} \bM_{\balpha,k}$, 
$\overline{\by}_{2k} = \tilde{\by}_k - \bU_{2k} \tilde{\bf}_k$, 
and 
$\bM_{\balpha,k} = \bU_{1k} (\bI_{\qgk} \otimes \bB_{k}^*)$.
Moreover, $\mathbf{H}_{\left[J_k,J_k\right]}$ indicates the $J$th row and the $J$th column in $\mathbf{H}_k$ for the sequence $J_k = \left\{r_k^2+1,r_k^2+2,\dots,\qgk r_k\right\}$.
The conditional posterior distribution of $\bbeta_k$, given $(\bY,\bs,\bSigma,\bgamma_k,r_k,\bA_k,\tilde{\bf}_k)$, is proportional to the multivariate Gaussian distribution $\mathcal{N}_{pr_r}(\bbeta_k | \bmuo_{\bbeta,k}, \Upsilono_{\bbeta,k})$, where $\Upsilono_{\bbeta,k} = (\Upsilonu_{\bbeta,k}^{-1} + \bM_{\bbeta,k}' \widetilde{\bSigma}_k^{-1} \bM_{\bbeta,k})^{-1}$ and $\bmuo_{\bbeta,k} = \Upsilono_{\bbeta,k} \bM_{\bbeta,k}' \widetilde{\bSigma}_k^{-1} \overline{\by}_{2k}$, with $\bM_{\bbeta,k} = \bU_{1k} (\bA_k \otimes \bI_p)$.

The posterior distribution of $\rho_k$, which is the probability of a response variable belonging to the low-rank group in state $k$, is the Beta distribution, $\mathcal{B}e(\rho_k | \overline{a}_\rho, \overline{b}_\rho)$ with $\overline{a}_\rho = \underline{a}_\rho + \qgk$, and $\overline{b}_\rho = \underline{b}_\rho + q - \qgk$.

Assuming that the initial distribution of $\bs$ is independent of $\bXi$, the rows $\bxi_k$ of $\bXi$ are independent a posteriori, and are drawn from $\bxi_k \sim \mathcal{D}ir(\overline{\bd})$, where
$\overline{\bd} = (d_{k1} + N_{k1}(\bs),\dots, d_{kK} + N_{kK}(\bs))$
and $N_{kl}(\bs) = |\{t : s_{t-1}\!=\!k, s_t \!= \!l\}|$ counts the numbers of transitions from state $k$ to state $l$ for the current draw of $\bs$.

Finally, to obtain the conditional posterior of the innovation covariance matrix, let us first rewrite $\check{\by}_t = (\by_t - \bV_{1s_t}' \bC_{s_t}' \bx_t - \bV_{2s_t}' \bf_{s_t}(\bx_t))$ and the model as
\begin{align*}
    \bW \check{\by}_t = \bD_t^{1/2} \be_{t}.
\end{align*}
Owing to the lower triangular structure of $\bW$, this system consists of a set of $j = 2,\dots,q$ equations, with equation $j$ having as dependent variable $\check{y}_{jt}$ and as independent variables $-\check{y}_{\ell t}$, with $\ell= 1,\dots,j-1$, with coefficients $w_{j\ell}$.
Therefore, multiplying the $j$th equation by $d_{jt}^{1/2}$ removes the heteroskedasticity associated with stochastic volatility.
Replicating this approach separately for each transformed equation $j$ leads to a Gaussian posterior distribution for the vector of coefficients of the $j$th equation, $\bw_j = (w_{j1},\dots,w_{jj-1})$'. We refer to \cite{cogley2005drifts} for further details.

The path of the stochastic volatility, $\bh_j = (h_{j1},\dots,h_{jT})'$, for each $j=1,\dots,q$, is drawn jointly from the approximate posterior distribution obtained using the mixture approach of \cite{omori2007stochastic}.
The posterior full conditional distribution of the initial state conditions $h_{j0}$ is conjugate and drawn from a Normal distribution with variance $\overline{\upsilon}_h^2 = 1/\underline{\upsilon}_h^2 + 1/\sigma_j^2$, and mean $\overline{\mu}_h = h_{j1}/(\overline{\upsilon}_h^2 \sigma_j^2)$ \citep{chan2018bayesian}.
The posterior full conditional distribution of $\sigma_j^2$ is conjugate $\sigma_j^2 | \bh_j \sim \mathcal{IG}(\overline{a}_\sigma,\overline{b}_\sigma)$, where $\overline{a}_\sigma = \underline{a}_\sigma + n$ and $\overline{b}_\sigma = \underline{b}_\sigma + \sum_{t=1}^T (h_{jt}-h_{jt-1})^2$.

\FloatBarrier
\section{Simulation study}
\label{sec:simulations}


In this section, we evaluate the performance of the proposed model in recovering the latent state chain, and the true group allocations of the response variables into the low-rank and the flexible classifications within each state. Then, we assess the effectiveness of the sampling procedure in estimating the rank and recovering the matrix of mean values of $\bY$, denoted by $\bM$.

We first describe in detail the data generating process (DGP) used for the different simulation studies.
The rows of the design matrix $\bX$ were sampled independently from $\mathcal{N}_p(\bzero,\bI_p)$, while the error terms, $\be_t$, were drawn from a multivariate normal distribution with a diagonal covariance matrix whose entries were uniformly chosen over the interval $(0.1,1)$.
The entries of the low-rank matrices $\bC_k$ were sampled uniformly from $(-3,-1.5)\cup(1.5,3)$. To enforce the reduced-rank structure, the smallest $p-r_k$ singular values of the matrix were set to zero.
The nonparametric term was generated as a sine wave with amplitude $2$, and the allocation of the response variables to the reduced-rank group was randomly assigned, given $\qgk$. 

We consider a total of six simulation settings, all of which use the configuration $p = q = 5$ and $ n = 100$, primarily differing in the composition of the hidden Markov chain and in the treatment of volatility in the model. The parameters in first two Scenarios (1 and 2) are $K = 2$, $\mathbf{\qg} = (4, 2)$, and $\br = (2, 1)$. In scenario 1, the chain $\bs$ is generated with a single regime switch at the 60\textsuperscript{th} observation of the time series, such that $s_1,\dots,s_{60} = 1$ and $s_{61},\dots,s_{100} = 2$. In scenario 2, $\bs$ is generated with random switches between the two regimes. The true DGP in scenario 3 considers $K = 1$ (no switch), $\qg = 3$, and $r=1$. In all three cases, estimation is performed under the assumption of constant volatility. The remaining three scenarios (4, 5, and 6) use the same parameter values as in the first scenarios, respectively, but estimation is conducted under a model that allows for stochastic volatility.

We run the proposed MS-PRR model for 50,000 MCMC iterations after discarding an additional 50,000 iterations as burn-in, and the hyperparameters are set to reflect noninformative priors. In detail,  
$\underline{a}_\rho = \underline{b}_\rho = 1$,
$\underline{a} = \underline{b} = 2$, 
$\nuu = q+1$, $\Psiu = \bI_q$,
and $\underline{\bd}$ is a $K$-dimensional vector of ones.

The point estimates of the binary allocation vectors $\hat{\bgamma}$, the ranks $\hat{\br}$, and the latent states $\hat{\bs}$ are obtained as their respective maximum a posteriori (MAP) estimates.
The performance of the estimator $\widehat{\bM}$ of the mean response matrix is assessed using the mean squared prediction error (MSPE), defined as \mbox{$\text{MSPE}=\|\widehat{\bM}-\bM\|_\text{F}^2/(Tq)$}, where $\widehat{\bM}$ is the posterior average of the predicted response matrix. Additionally, we compute the mean squared error (MSE) of the fitted values $\widehat{\bY}$, defined as \mbox{$\text{MSE}=\|\widehat{\bY}-\bY\|_\text{F}^2/(Tq)$}.
To evaluate the classification performance of the state chain, $\bs$, and the allocation parameter, $\bgamma$, we use accuracy and the $F_1$ score as metrics, both ranging from 0 to 1. Higher values indicate better agreement between the estimated and true response groupings, with 1 indicating a perfect classification.

Tables \ref{tab:sims_SV0} and \ref{tab:sims_SV1} summarise the simulation results by providing the average MSE and MSPE over 20 independent experiments of each setting, the average of the estimated number of low-rank responses, $\hat{q}_{\gamma,k}$, the estimated rank, $\hat{r}_k$, the accuracies and the $F_1$ scores.
The proposed method accurately recovers the true latent state sequence, as indicated by the low mean squared error. The allocation vectors are also accurately estimated, with accuracy and $F_1$ scores consistently attaining values close to 1. Additionally, the estimated ranks closely match the true ranks on average, demonstrating the effectiveness of the approach in recovering the underlying model structure.

\begin{table}[!th]
\centering
\resizebox{\columnwidth}{!}{%
\begin{tabular}{@{}c|c|cccccccc@{}}
  Scenario &  Model   & MSE    & MSPE  & $\hat{\mathbf{q}}_{\bgamma}$ & $\hat{\br}$   & Accuracy  ($\bgamma$) & $F_1$ score ($\bgamma$) & Accuracy ($\bs$)  & $F_1$ score ($\bs$) \\ 
  \midrule
  \multirow{4}{*}{1} & MS-PRR & \textbf{0.682}  & \textbf{0.381} & (4.000,2.000) & (2.100,1.050) & (1.000,0.900) & (1.000,0.897) & 0.996   & 0.996 \\
  & MS-PRR-NO-GP  & 1.291  & 0.897 & (4.000,4.000) & (2.050,1.000) & (1.000,0.580) & (1.000,0.650) & 1.000  & 1.000  \\
  & PRR-GP  & 5.600  & 5.279 & 3.611   & 2.200 & (0.780,0.520)  & (0.841,0.563)  & \cellcolor[HTML]{EFEFEF} & \cellcolor[HTML]{EFEFEF} \\
  & PRR  & 13.452 & 6.587 & 2.889   & 1.800   & (0.600,0.460) & (0.691,0.472) & \cellcolor[HTML]{EFEFEF} & \cellcolor[HTML]{EFEFEF} \\
  \hline
\multirow{4}{*}{2} & MS-PRR & \textbf{0.588}  & \textbf{0.309} & (4.000,2.263) & (2.000,1.000) & (1.000,1.000) & (1.000,1.000)  & 0.999  & 0.999  \\
  & MS-PRR-NO-GP  & 1.345  & 0.960 & (4.000,4.000)  & (1.389,1.611) & (0.756,0.422) & (0.847,0.519)  & 0.997   & 0.997   \\
  & PRR-GP    & 5.779  & 5.499 & 3.421  & 2.222   & (0.767,0.478)  & (0.839,0.528)  & \cellcolor[HTML]{EFEFEF} & \cellcolor[HTML]{EFEFEF} \\
  & PRR   & 16.773 & 6.864 & 2.895 & 1.556   & (0.533,0.511)  & (0.641,0.485)   & \cellcolor[HTML]{EFEFEF} & \cellcolor[HTML]{EFEFEF} \\
  \hline
\multirow{4}{*}{3} & MS-PRR & \textbf{0.583}  & \textbf{0.284} & 3.000   & 1.000  & 1.000  & 1.000  & 1.000  & 1.000  \\
  & MS-PRR-NO-GP    & 1.478  & 0.958 & 3.889  & 1.000 & 0.689  & 0.762  & 1.000 & 1.000  \\
  & PRR-GP    & 0.585  & 0.296 & 3.000  & 1.000   & 1.000  & 1.000 & \cellcolor[HTML]{EFEFEF} & \cellcolor[HTML]{EFEFEF} \\
  & PRR       & 8.967  & 0.785 & 4.000 & 1.000  & 0.800  & 0.857   & \cellcolor[HTML]{EFEFEF} & \cellcolor[HTML]{EFEFEF} \\ \bottomrule
\end{tabular}}
\caption{Average metrics across 20 replications of each scenario where the estimation procedure assumes constant volatility. For each competitor we report the mean squared error (MSE), the mean squared prediction error (MSPE), the estimated number of low-rank responses ($\hat{\mathbf{q}}_{\bgamma}$), the estimated rank ($\hat{\br}$), the accuracies and the $F_1$ scores of $\bgamma$ and $\bs$. The smallest errors are in boldface. The entries of $\hat{q}_{\bgamma}$, $\hat{\br}$, accuracy and F1 score of $\bgamma$ corresponding to the Markov-switching models in scenarios 1 and 2 report the estimates for each state; the models without Markov-switching have only one estimate of $\hat{\mathbf{q}}_{\bgamma}=\hat{q}_{\bgamma,1}$ and $\hat{\br}=\hat{r}_{1}$, but report the accuracy and F1 score relative to the true $\bgamma = (\bgamma_1,\bgamma_2)$.}
\label{tab:sims_SV0}
\end{table}

\begin{table}[!th]
\centering
\resizebox{\columnwidth}{!}{%
\begin{tabular}{@{}c|c|cccccccc@{}}
Scenario & Model & MSE & MSPE & $\hat{\mathbf{q}}_{\bgamma}$ & $\hat{\br}$ & Accuracy ($\bgamma$) & $F_1$ score ($\bgamma$) & Accuracy ($\bs$) & $F_1$ score ($\bs$) \\ \midrule
\multirow{4}{*}{4} & MS-PRR & \textbf{0.555} & \textbf{0.282} & (4.000,2.000) & (2.050,1.000) & (1.000,0.980) & (1.000,0.975) & 0.999 & 0.999 \\
 & MS-PRR-NO-GP & 1.503 & 1.124 & (4.000,3.650) & (2.050,1.000) & (1.000,0.670) & (1.000,0.713) & 1.000 & 1.000 \\
 & PRR-GP & 9.456 & 9.086 & 2.600 & 1.250 & (0.640,0.620) & (0.718,0.600) & \cellcolor[HTML]{EFEFEF} & \cellcolor[HTML]{EFEFEF} \\
 & PRR & 22.798 & 9.468 & 2.050 & 1.000 & (0.210,0.370) & (0.345,0.220) & \cellcolor[HTML]{EFEFEF} & \cellcolor[HTML]{EFEFEF} \\
\midrule
\multirow{4}{*}{5} & MS-PRR & \textbf{0.545} & \textbf{0.284} & (4.000,2.000) & (2.050,1.000) & (1.000,1.000) & (1.000,1.000) & 0.998 & 0.998 \\
 & MS-PRR-NO-GP & 1.355 & 0.994 & (4.000,3.800) & (1.500,1.500) & (0.780,0.500) & (0.863,0.580) & 0.997 & 0.997 \\
 & PRR-GP & 8.302 & 7.982 & 2.850 & 1.350 & (0.610,0.570) & (0.697,0.557) & \cellcolor[HTML]{EFEFEF} & \cellcolor[HTML]{EFEFEF} \\
 & PRR & 16.015 & 7.674 & 2.200 & 1.050 & (0.360,0.520) & (0.475,0.425) & \cellcolor[HTML]{EFEFEF} & \cellcolor[HTML]{EFEFEF} \\ \midrule
\multirow{4}{*}{6} & MS-PRR & \textbf{0.633} & \textbf{0.343} & 2.909 & 1.091 & 0.982 & 0.982 & 0.998 & 1.000 \\
 & MS-PRR-NO-GP & 1.440 & 0.942 & 4.000 & 1.000 & 0.727 & 0.805 & 0.997 & 1.000 \\
 & PRR-GP & 0.636 & 0.344 & 2.909 & 1.091 & 0.982 & 0.982 & \cellcolor[HTML]{EFEFEF} & \cellcolor[HTML]{EFEFEF} \\
 & PRR & 11.769 & 0.794 & 3.909 & 1.000 & 0.818 & 0.870 & \cellcolor[HTML]{EFEFEF} & \cellcolor[HTML]{EFEFEF} \\ \bottomrule
\end{tabular}}
\caption{Average metrics across 20 replications of each scenario where the estimation procedure assumes stochastic volatility. For each competitor we report the mean squared error (MSE), the mean squared prediction error (MSPE), the estimated number of low-rank responses ($\hat{\mathbf{q}}_{\bgamma}$), the estimated rank ($\hat{\br}$), the accuracies and the $F_1$ scores of $\bgamma$ and $\bs$. The smallest errors are in boldface. The entries of $\hat{q}_{\bgamma}$, $\hat{\br}$, accuracy and F1 score of $\bgamma$ corresponding to the Markov-switching models in scenarios 1 and 2 report the estimates for each state; the models without Markov-switching have only one estimate of $\hat{\mathbf{q}}_{\bgamma}=\hat{q}_{\bgamma,1}$ and $\hat{\br}=\hat{r}_{1}$, but report the accuracy and F1 score relative to the true $\bgamma = (\bgamma_1,\bgamma_2)$.}
\label{tab:sims_SV1}
\end{table}

In all scenarios, adding the GP prior allows for a better estimation in both models with and without Markov switching. An analogous conclusion is observable when comparing the models with Markov switching to their counterparts. Overall, MS-PRR, combining both a GP prior and Markov-switching, achieves the best performance among the four models. MS-PRR-NO-GP without using the GP specification oversimplifies the model by allocating, on average, approximately four responses to the low-rank group(s). 
However, it is the second best model in the settings where $K=2$. The PRR-GP, which is the PRR model with a GP component but no Markov switching component, outperforms the competitors when $K=1$, even though MS-PRR is estimating a one-state chain as the true DGP. Nonetheless, the proposed approach still obtains the lowest error by accurately estimating one state, and adding more flexibility through the GP specification. Finally, standard PRR model exhibits the weakest performance throughout all scenarios.

\begin{figure}[t!h]
\centering
    \begin{minipage}{0.6\linewidth}
        \includegraphics[width=\linewidth]{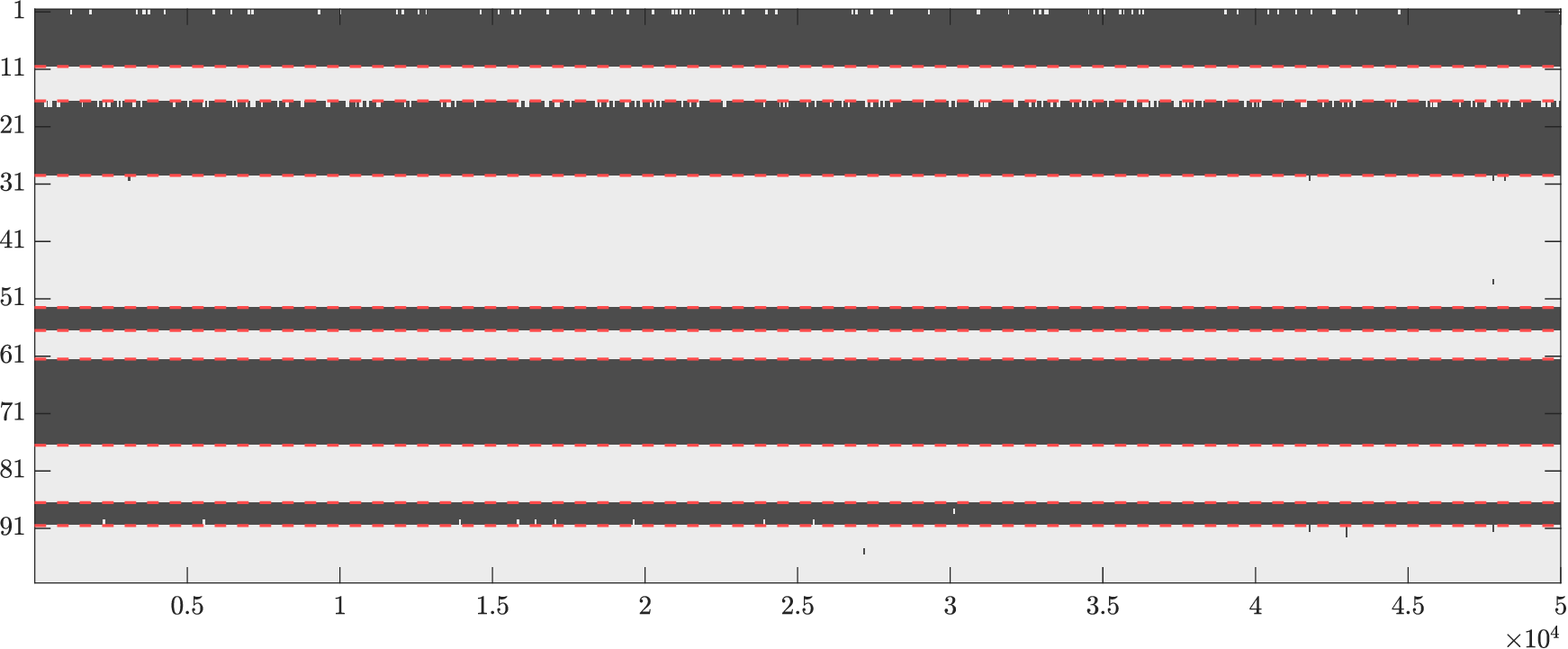} \\
        \centering \footnotesize(a)
    \end{minipage}
    \hfill
    \begin{minipage}{0.37\linewidth}
        \includegraphics[width=\linewidth]{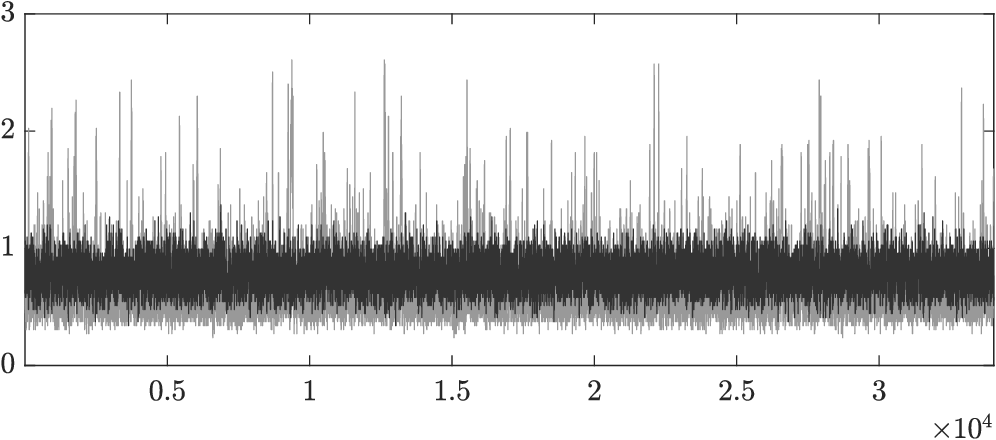} \\ 
        \includegraphics[width=\linewidth]{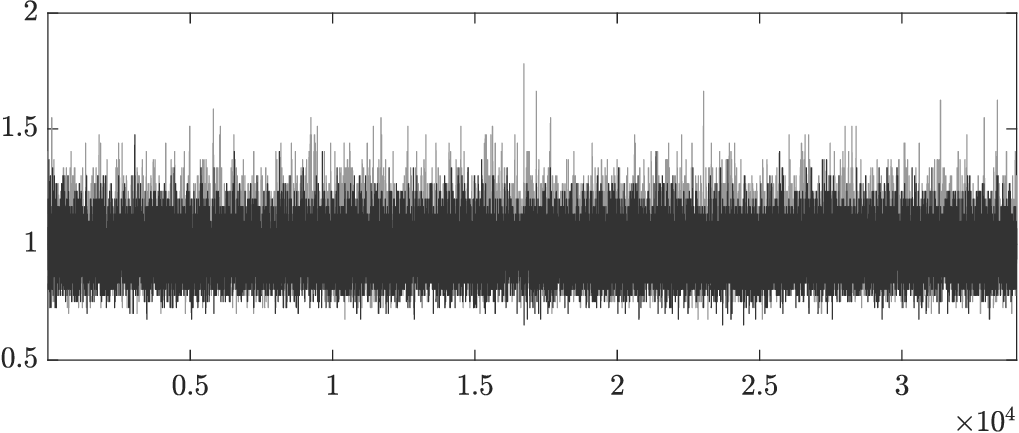} \\ 
        \centering \footnotesize(b)
    \end{minipage}
\caption{Panel (a): trace plot of $\bs$ in scenario 2. Each colour represents one state, while a dashed horizontal red line indicates the time when a true regime switch occurs. Panel (b): trace plots of the Gaussian process hyperparameters $\sigma_{f,k}^2$ (top) and $\zeta_k$ (bottom) in scenario 2.}
\label{fig:2_traces_sig2fz}
\end{figure}

The results discussed above are supported by the trace plots and posterior distributions of the parameters of interest.
Panel (a) of Figure~\ref{fig:2_traces_sig2fz} shows the accurate recovery of the hidden Markov chain, with a distinct separation between states, even in the presence of random switching.
When looking at the hyperparameters of the Gaussian process (panel (b)), $\sigma_{f,k}^2$ and $\zeta_{k}$ reach the convergence with overlaps in the support across the two states.

\begin{figure}[t!h]
\centering
\begin{tabular}{cc}
    \includegraphics[width=0.72\linewidth]{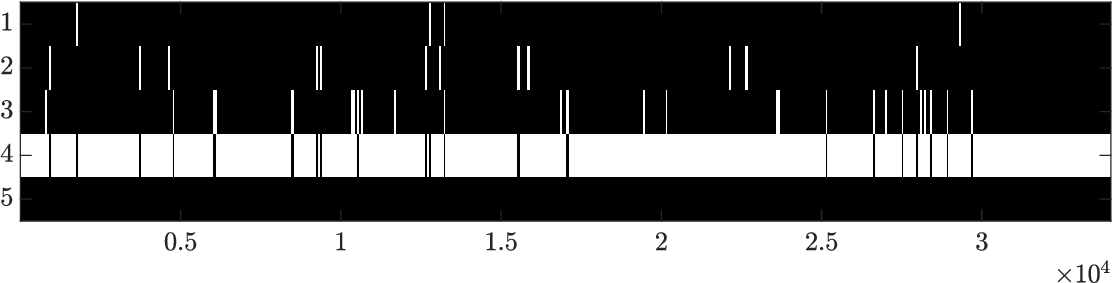} & 
    \includegraphics[width=0.2\linewidth]{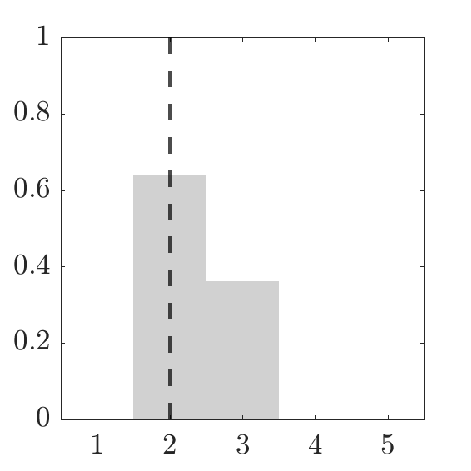} \\
    \includegraphics[width=0.72\linewidth]{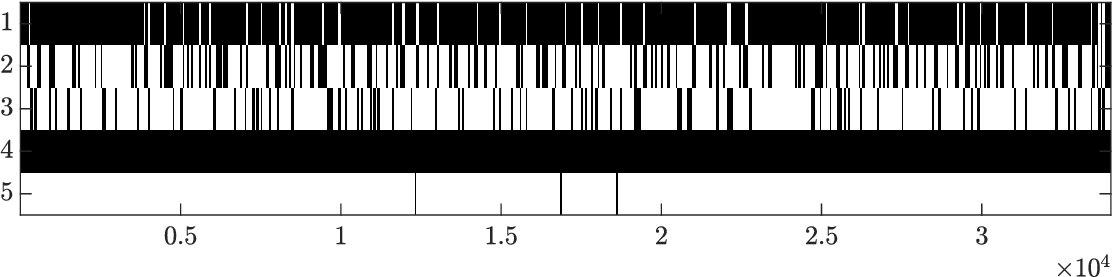} & 
    \includegraphics[width=0.2\linewidth]{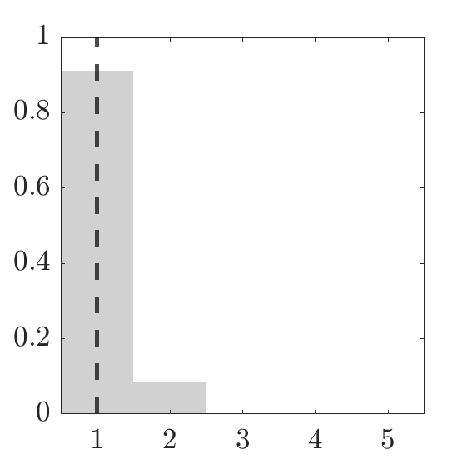}
\end{tabular}
\caption{Trace plots of $\bgamma$ and posterior distribution of $\br$ in scenario 2. The top row refers to state 1, where the true allocation vector is $\bgamma_{1} = (1,1,1,0,1)$. The second row to state 2, where the true allocation vector is $\bgamma_{2} = (1,0,0,1,0)$. The true values of the rank are indicated with dashed vertical lines.}
\label{fig:2_traceg_postr}
\end{figure}

\begin{figure}[t!h]
\centering
\begin{tabular}{ccccc}
    \includegraphics[width=0.18\linewidth]{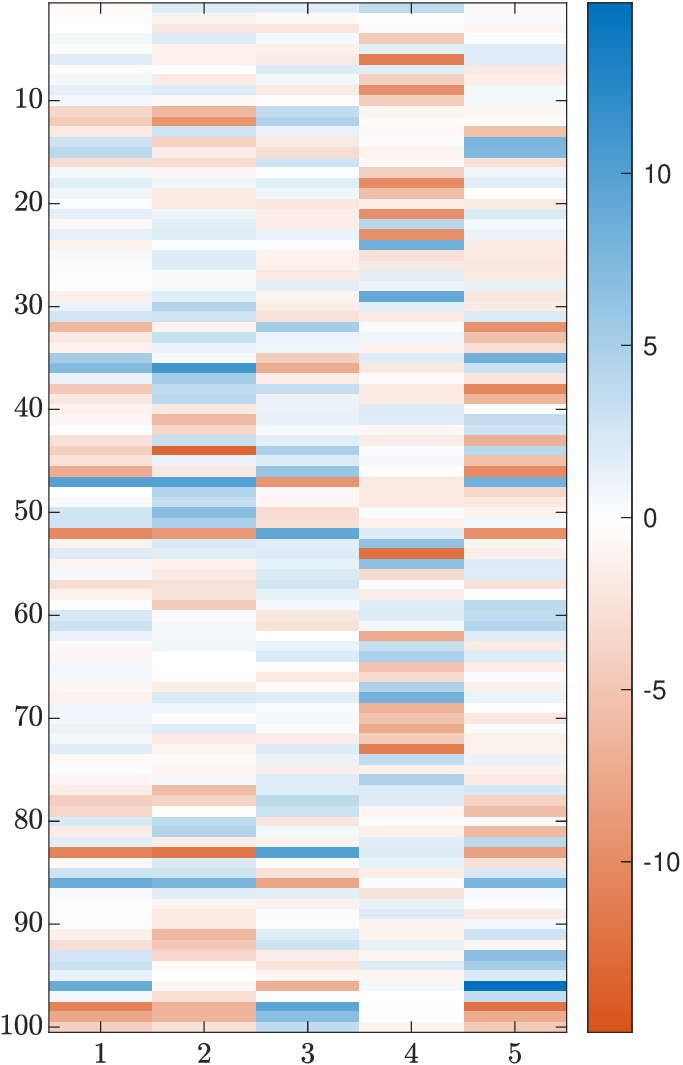} &
    \includegraphics[width=0.18\linewidth]{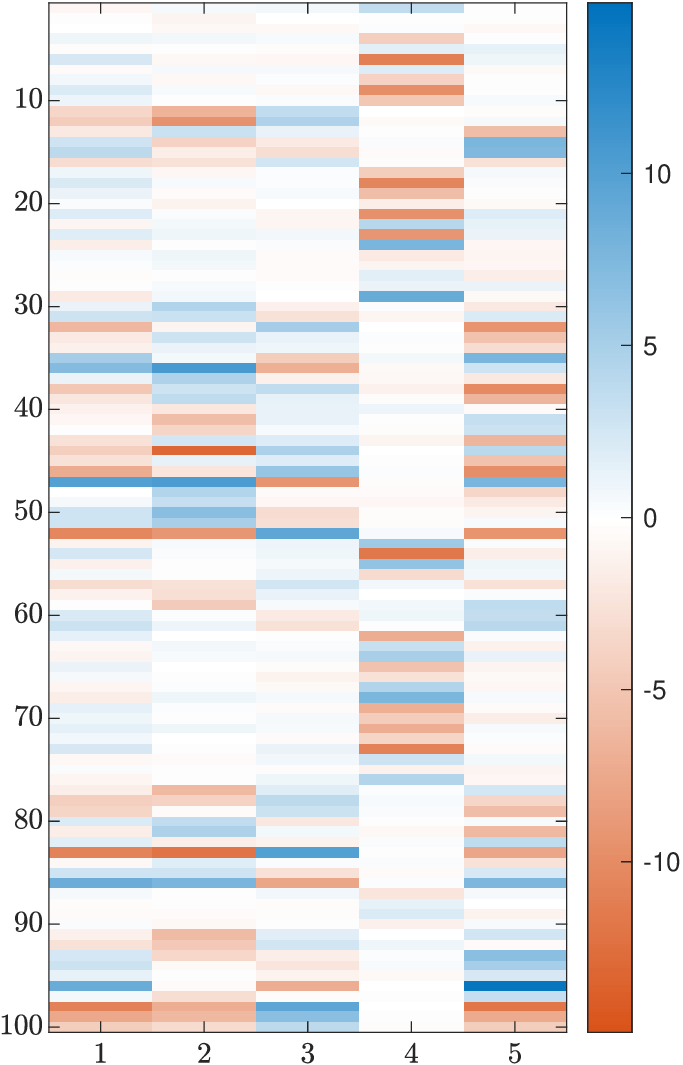} &
    \includegraphics[width=0.18\linewidth]{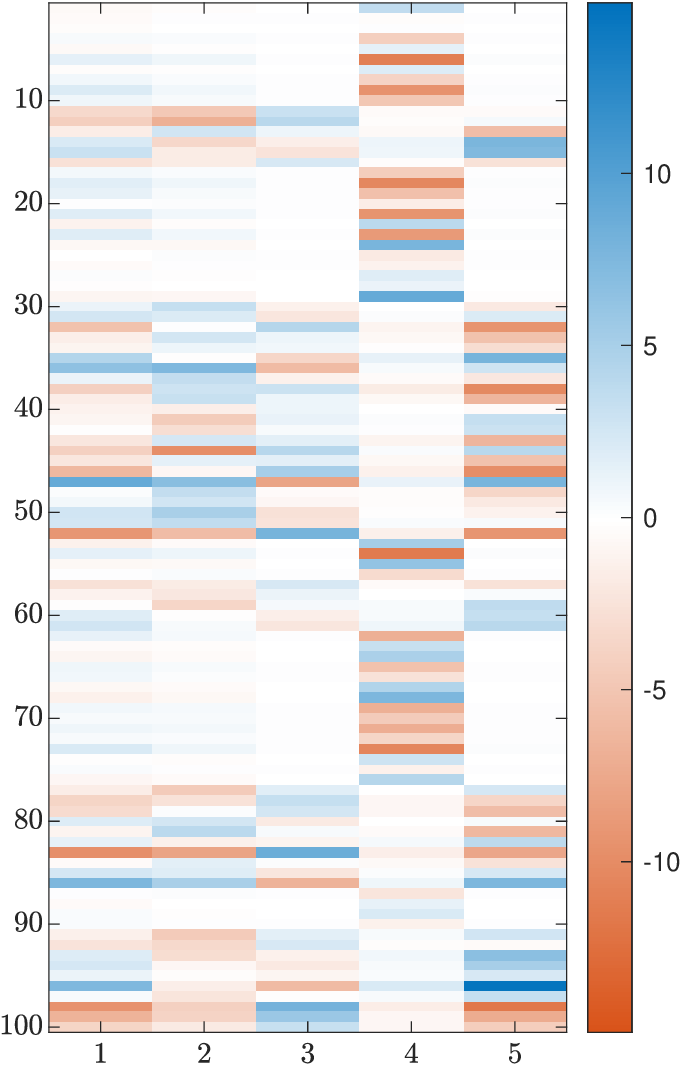} &
    \includegraphics[width=0.18\linewidth]{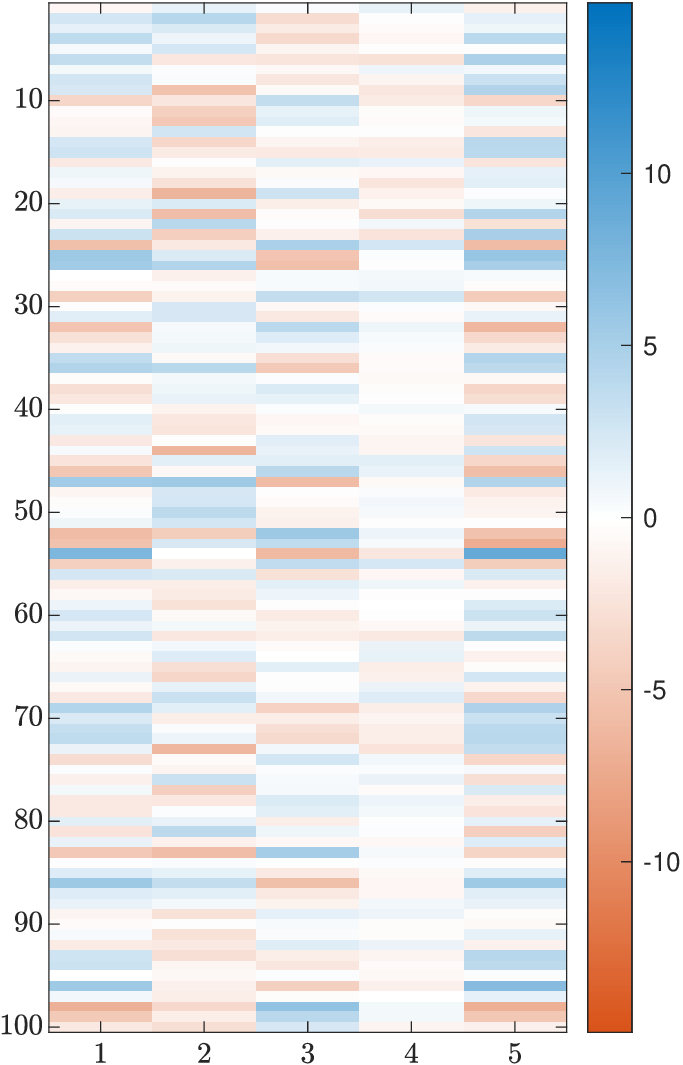} &
    \includegraphics[width=0.18\linewidth]{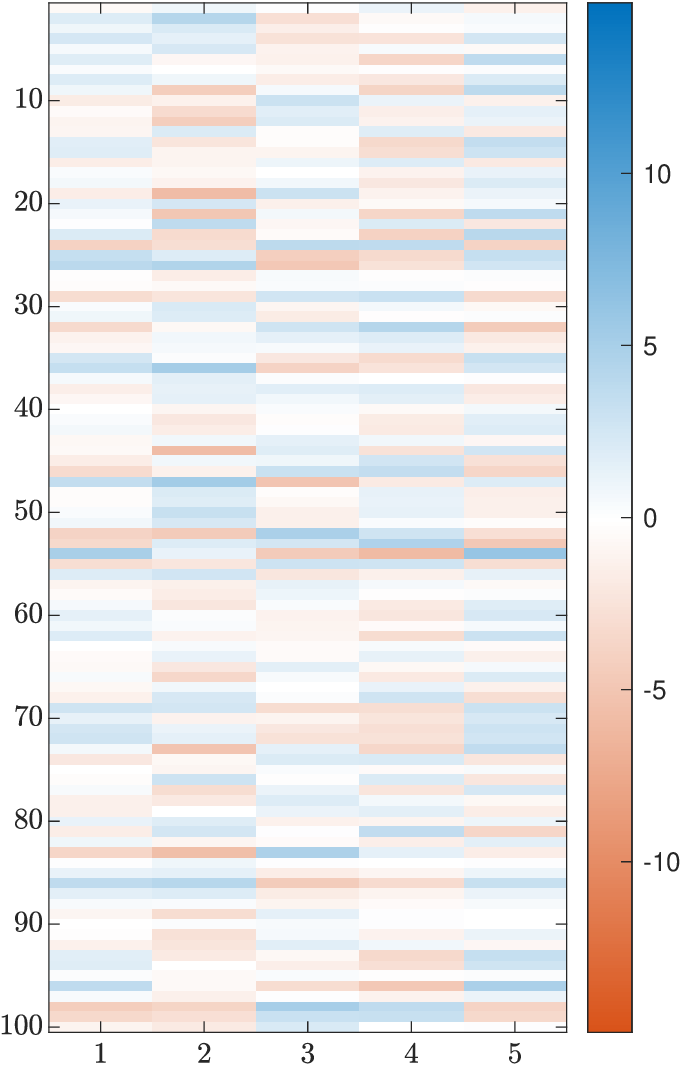}
    \\ \vspace{-1ex}
    \footnotesize (a) $M_0$ &
    \footnotesize (b) MS-PRR &
    \footnotesize (c) MS-PRR-NO-GP &
    \footnotesize (d) PRR-GP &
    \footnotesize (e) PRR
\end{tabular}
\caption{True mean matrix (a), and estimated mean matrix (b-e) in setting 2 under the different models.}
\label{fig:2_M}
\end{figure}

The model also successfully identifies the true response groupings and their associated ranks (Figure \ref{fig:2_traceg_postr}), and it also obtains a close estimation to the true mean matrix, as illustrated in Figure~\ref{fig:2_M}.
Overall, these results support the robustness and effectiveness of the proposed approach across varying conditions.

\section{Empirical applications}
\label{sec:app}

In this section, we apply the proposed method to detect switches among regimes on two different applications. 
The first application relies on US macroeconomic data, while the second one focuses on the energy market. 

\subsection{Macroeconomic data}

Quarterly US macroeconomic data generally show presence of possible switches among states, thus requiring the use of Markov-switching model. 
Following \cite{pintado2025bayesian}, which employ a partial reduced-rank model, we rely on the same dataset covering the period from 2000 to 2023 to detect possible recessionary or expansionary periods.
%
The $q=5$ responses are the index of industrial production ($y_1$), personal consumption of food and drinks ($y_2$), unemployment rate ($y_3$), volume index of imports of goods and services ($y_4$), and volume index of exports of goods and services ($y_5$).
The $p=5$ covariates are civilian labour force level ($x_1$), median weekly earnings ($x_2$), price index of imports of goods and services ($x_3$), price index of exports of goods and services ($x_4$), and price index of final consumption expenditure ($x_5$) 
(see the Supplement for details).
All variables were standardised before conducting the analysis.
In \cite{pintado2025bayesian}, the authors explored whether the years before and after the pandemic outbreak exhibit similar drivers and structure. 
They found a substantial shift in the low-rank structure of the coefficient matrix, transitioning from a nearly fully reduced-rank form in the pre-COVID era to a structure approaching full rank in the subsequent period. 
Moreover, a change in the regression structure was observed over time.


Building upon these findings, we now apply the MS-PRR model 
to investigate how the response allocation vector evolves over time and quantify the associated uncertainty.
Specifically, we seek to capture temporal changes in the regression structure as regime switches in a hidden Markov chain. 
We set the number of regimes to $K=2$ to detect possible recession or expansion period and in particular to identify if any changes appear before and after the COVID-19 period as suggested in \cite{pintado2025bayesian}.

\begin{figure}[t!h]
\centering
    \includegraphics[width=0.8\linewidth]{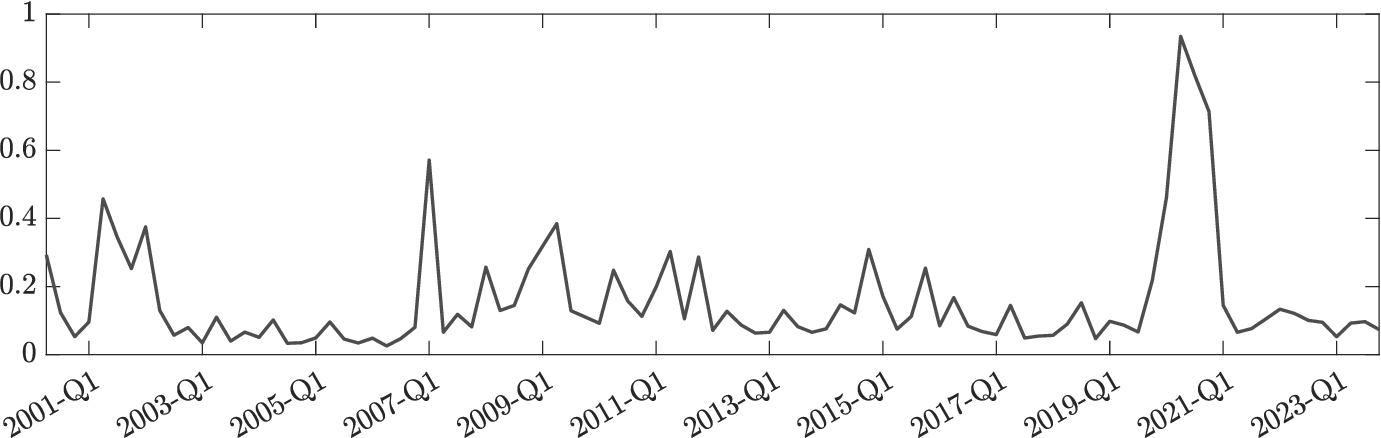}
\caption{Posterior probability of each time point belonging to state 2.}
\label{fig:macro_s}
\end{figure}

Figure~\ref{fig:macro_s} reveals that our method has identified one main regime switch corresponding to the COVID pandemic.
Moreover, we identify two other periods with high probability of belonging to the second state, which are the second quarter of 2001 (0.4575) and the first quarter of 2007 (0.5712).
The latter points towards the start of the 2008 financial crisis, while the former aligns with the early 2000s recession in the United States.
The estimated allocations are $\hat{\bgamma}_1 = (0,1,0,1,1)$ and $\hat{\bgamma}_2 = (1,0,1,1,1)$. The first state observes a rank-1 coefficient matrix for the low-rank group, while a rank of 3 was estimated for the second state. 

\begin{figure}[t!h]
\centering
\begin{tabular}{cc}    
    \includegraphics[width=0.4\linewidth]{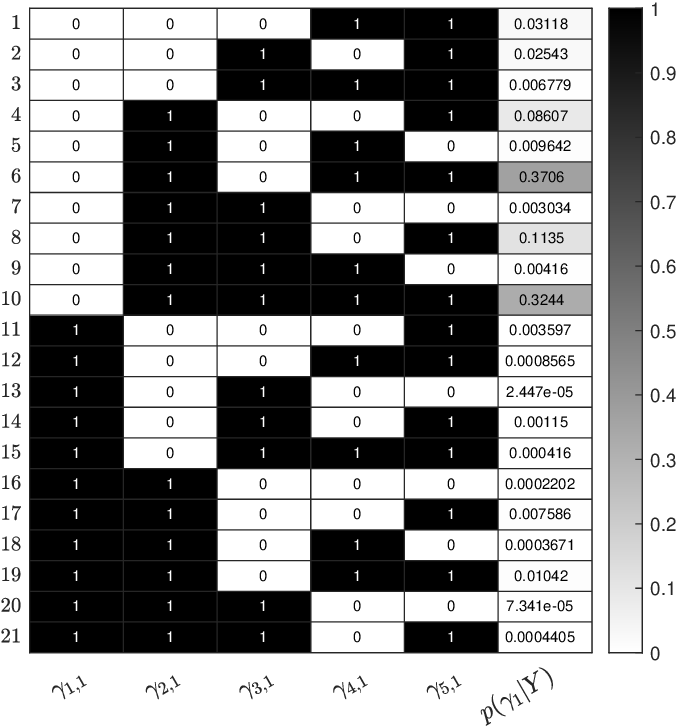} & 
    \includegraphics[width=0.4\linewidth]{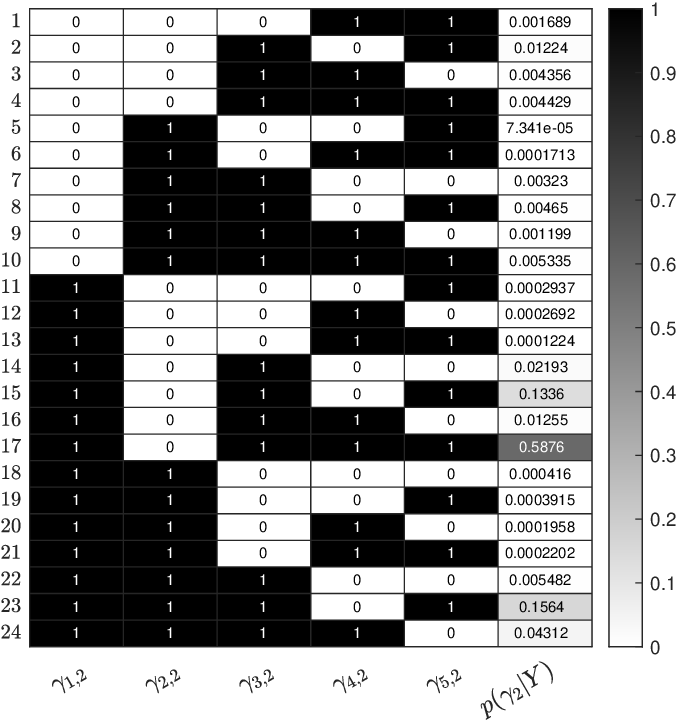}
    \end{tabular}
\caption{Posterior distribution of the allocation vector, $\bgamma_k$, for the first state (left) and the second state (right).}
\label{fig:macro_postg}
\end{figure}

In Figure~\ref{fig:macro_postg}, we notice that the first state exhibits greater uncertainty in the allocation structure, with its posterior mass distributed almost equally among two main candidate vectors, which differ in the inclusion or exclusion of $y_3$ (unemployment rate) in the low-rank group.
In contrast, the second state, which resembles the crisis period, exhibits a strong posterior concentration around its mode, which includes industrial production, unemployment rate and import/export volume index.

\begin{figure}[t!h]
\centering
\begin{tabular}{cc}
     \includegraphics[width=0.20\linewidth]{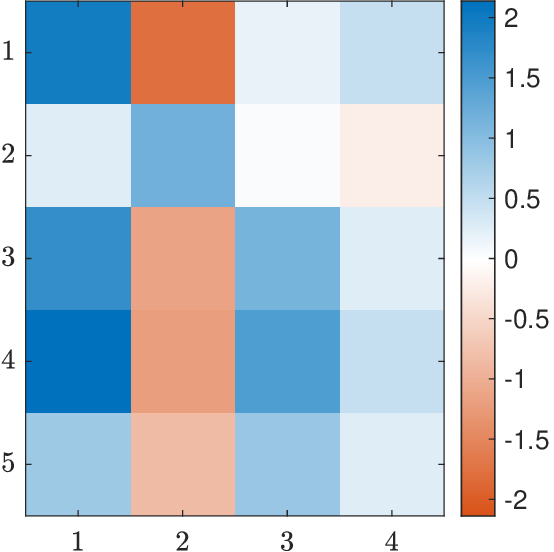} \hspace{1cm}
     & \includegraphics[width=0.20\linewidth]{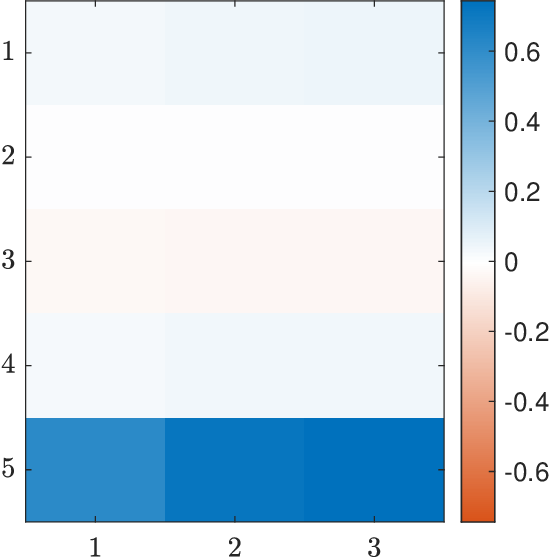}
\end{tabular}
\caption{ Estimated coefficient matrix for state 1 (left) and state 2 (right), where responses are labelled on the horizontal axis and covariates on the vertical axis.}
\label{fig:macro_M}
\end{figure}

Figure~\ref{fig:macro_M} provides the estimated coefficient matrix according to the two states. The pattern changes among the two regimes, where the first regime has variation among positive and negative values, highlighting an accurate signal of the covariates. In period of crisis, as provided by the right panel, we notice a weak signal of the covariates and a concentration in the last component. 

To wrap up, the MS-PRR model was able to identify recession periods and to reveal how economic relations change over time and regimes.

\subsection{Commodity sector}

The commodity market and in particular the energy sector is often characterised by complex interdependencies and dynamic interactions among energy prices and macroeconomic conditions, where structural changes may unfold over time. 
Understanding these relationships is essential for modelling the evolution of energy markets, which are potentially affected by periods of geopolitical and economic shifts such the current ongoing Russian invasion of Ukraine or the previous oil war in Iraq.
To further illustrate the practical value of our proposed MS-PRR approach, we rely on a multivariate quarterly commodity dataset spanning from January 1997 to April 2022. 
The response variables considered refer to the classical commodity prices, which are important for the global economy market: oil price ($y_1$), gas price ($y_2$), wholesale electricity ($y_3$), industrial $\text{CO}_2$ price ($y_4$), coal price ($y_5$).
As covariate, we include different macroeconomic indicators that are strongly influenced by changes in the commodity market: industrial production index ($x_1$), producer price index ($x_2$), inflation rate ($x_3$), volatility index ($x_4$), and consumer confidence index ($x_5$).

\begin{figure}[t!h]
\centering
    \includegraphics[width=0.8\linewidth]{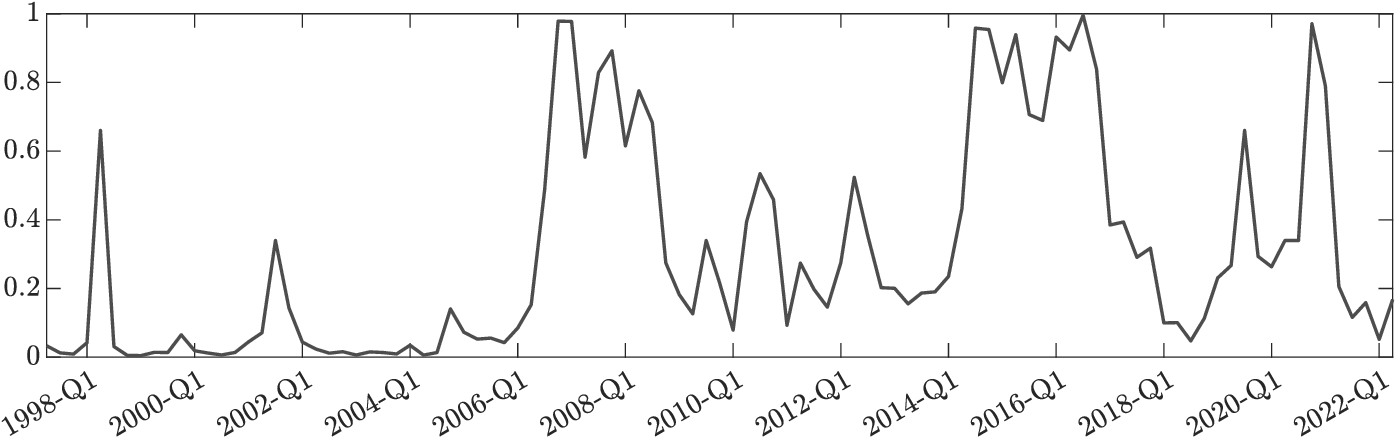}
\caption{Posterior probability of each time point belonging to state 2.}
\label{fig:energy_s}
\end{figure}

As in the previous experiment, we rely on two regimes and Figure~\ref{fig:energy_s} reports the posterior probability over time of belonging to the second state.
We identify three main switches that last for different periods: (i) the 2008 oil price boom and the financial crisis, when the oil prices reached the highest record; (ii) the 2014--2016 oil price collapse; (iii) the current energy crisis related to the Russian invasion of Ukraine.
The proposed model identifies also other lower-probability switches associated to Organization of the Petroleum Exporting Countries (OPEC) production cuts in late 1990s and the 2010-2013 European sovereign debt crisis. 
Thus, MS-PRR model was able to successfully detect both severe and moderate disruptions in the commodity markets.

\begin{figure}[t!h]
\centering
\begin{tabular}{cc}
    \includegraphics[width=0.72\linewidth]{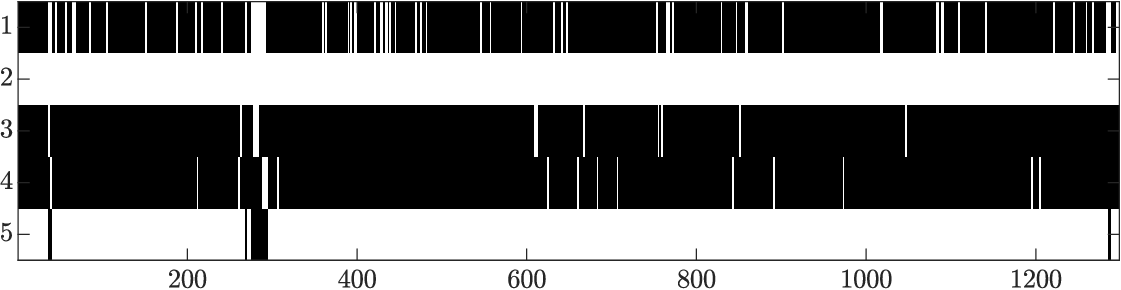} & 
    \includegraphics[width=0.2\linewidth]{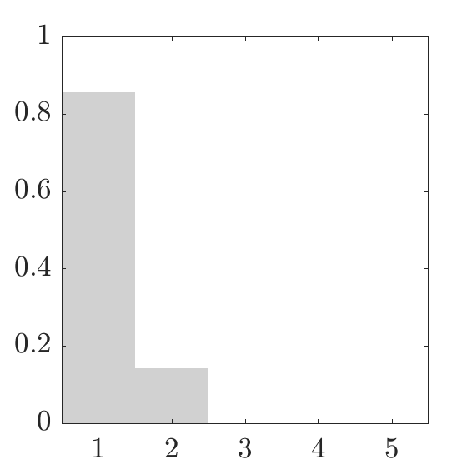} \\
    \includegraphics[width=0.72\linewidth]{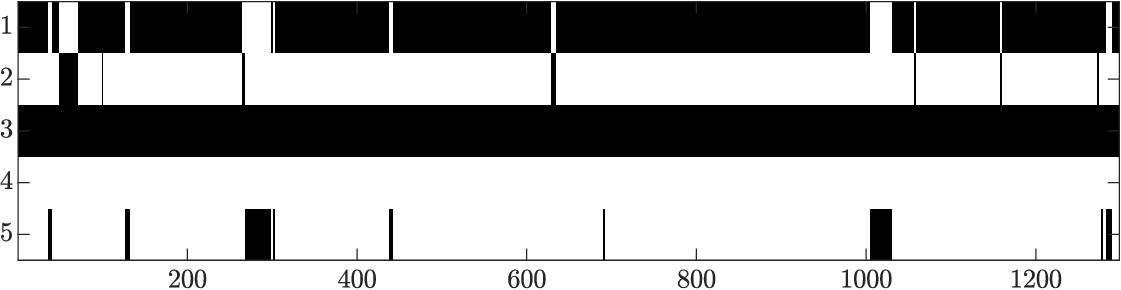} & 
    \includegraphics[width=0.2\linewidth]{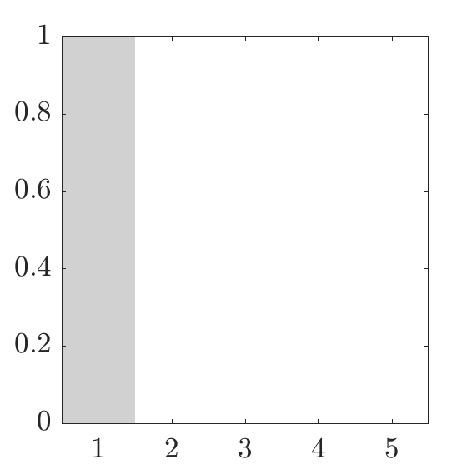}
\end{tabular}
\caption{Trace plots of $\bgamma$ and posterior distribution of $\br$ after 50,000 burn-in iterations. The top row is state 1, and the second row is state 2.}
\label{fig:energy_traceg_postr}
\end{figure}

In Figure~\ref{fig:energy_traceg_postr}, the trace plots of $\bgamma_k$ and the posterior distribution of the rank was provided.
The first state includes oil ($y_1$), wholesale electricity ($y_3$), and industrial $\text{CO}_2$ ($y_4$) prices in the group with a lower-rank coefficient matrix, thus highlighting the comovements among these three commodities which is consistent with integrated energy and industrial input markets in calm periods. 
Indeed, in ``stable'' or calm period, these three response variables have linear relationship with macroeconomic variables as expected. 
The second state, however, excludes the fourth response since the relationship between industrial $\text{CO}_2$ prices and macroeconomic indicators becomes nonlinear explained by the nonparametric GP component and more volatile during period of turmoil, revealing a structural change in how industrial $\text{CO}_2$ prices relate to the covariates. 
On the other hand, the other two responses, electricity and oil, maintain a simpler co-movement even in period of crisis  due to fuel substitution and common demand drivers. 
By looking at the rank posterior distribution, both states have an estimated rank equal to $1$, but the structure of the low-rank group changes.


\begin{figure}[t!h]
\centering
\begin{tabular}{c}
    \includegraphics[width=0.8\linewidth]{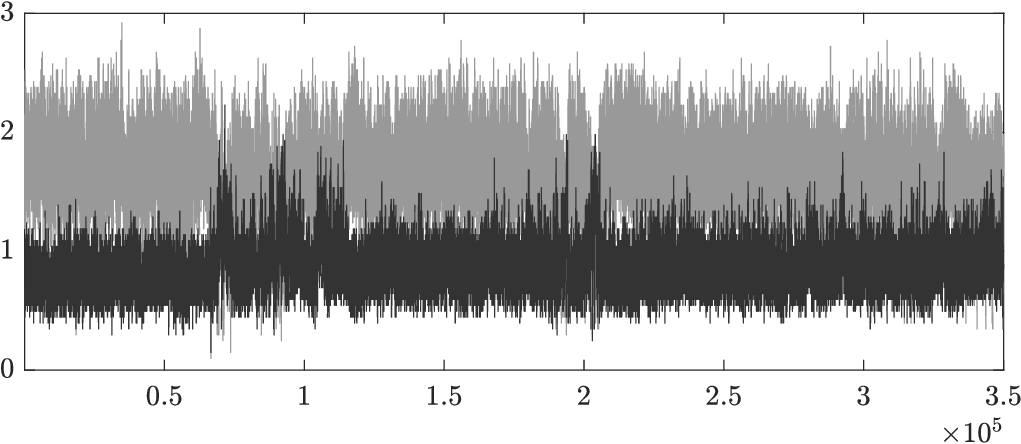} \\
    \includegraphics[width=0.8\linewidth]{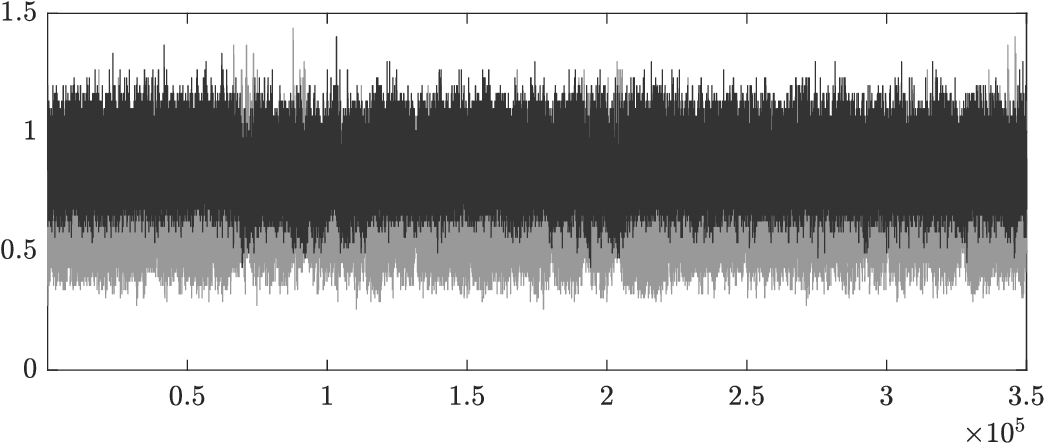}
\end{tabular}
\caption{Trace plots of the Gaussian process hyperparameters $\sigma_f^2$ (top) and $\zeta$ (bottom) for state 1 (light gray) and state 2 (dark gray) after 50,000 burn-in iterations.}
\label{fig:energy_si2f_zeta}
\end{figure}

Figure~\ref{fig:energy_si2f_zeta} illustrates the trace plots of the hyperparameters of the Gaussian process for each state and the complexity and smoothness of the nonlinear component in the MS-PRR model.
The level of separation of the posterior draws of $\sigma^2_f$ suggests the model can clearly distinguish the two states based on the characteristics of the nonlinear component. 
This result is also confirmed by the draws of the scale $\zeta$.

In conclusion, the proposed model was able to detect major regime switches in the energy markets and highlight the role of industrial $\text{CO}_2$ price as a non-linear, sensitive to crisis component.

\section{Concluding remarks}
\label{sec:conc}

We have proposed MS-PRR, a novel Bayesian extension of the partial reduced-rank regression model incorporating both time-varying parameters through a Markov-switching mechanism and further flexibility offered by a Gaussian process on complex terms of the regression. 
Our approach combines the interpretability and parsimony of reduced-rank regression with the adaptability of Gaussian processes for modelling complex dependencies, and the dynamic capabilities of Hidden Markov Models for detecting regime changes, particularly in the response clustering. The result is a flexible yet structured framework for modelling multivariate time series with evolving relationships among the responses and the covariates and by considering time-varying volatility.

Through simulation studies, we demonstrated the model's effectiveness in recovering both latent regime transitions and the associated response grouping structures, while providing uncertainty quantification about these estimations. In the two empirical applications to macroeconomic and commodity data, the proposed method successfully identified periods of changes corresponding to major economic, financial and energy crises.

Overall, the MS-PRR framework is an interpretable approach towards intricate structures in reduced-rank regression in terms of potential groups in the responses and time-varying dynamics, offering both methodological innovations and empirical relevance.
Further extensions of the model may consider more scalable inference methods for large dimensional datasets or generalisations to high-frequency data settings, which increases the domain of applications to other financial data or medicine, where the granularity of the data is often daily or even per second.


\bibliographystyle{apalike}
\bibliography{biblio}

\end{document}